\documentclass[prl,aps,
twocolumn,
nofootinbib,
floatfix,
superscriptaddress]{revtex4}

\usepackage{amsmath,amssymb}
\usepackage{mathrsfs}
\usepackage{color}
\usepackage{centernot}
\usepackage{tikz, pict2e}
\usepackage{lmodern}
\usepackage{tkz-euclide}
\usepackage[utf8]{inputenc}

\usetikzlibrary{calc,backgrounds, intersections}
\usetikzlibrary{arrows,shapes,positioning,shadows,trees}
\usetikzlibrary{mindmap}
\usetikzlibrary{decorations.pathreplacing}
\usetikzlibrary{decorations.pathmorphing}
\usetikzlibrary{decorations.markings}
\usetikzlibrary{matrix}

\tikzstyle arrowstyle=[scale=1]
\tikzstyle directed=[postaction={decorate,decoration={markings,
    mark=at position .5 with {\arrow[arrowstyle]{stealth}}}}]
\tikzstyle reverse directed=[postaction={decorate,decoration={markings,
    mark=at position .5 with {\arrowreversed[arrowstyle]{stealth};}}}]

\definecolor{redb}{rgb}{0.700, 0.000, 0.300}
\DeclareMathAlphabet\mathbfcal{OMS}{cmsy}{b}{n}

\def\r{\rho}

\def\aa1{\phi}
\def\cc1{\psi}

\usepackage{hyperref}

\DeclareRobustCommand{\cross}[1]{%
\begin{tikzpicture}[cross/.style={cross out, draw, minimum size=.6em, inner sep=0}]
   \node[very thick, cross=4pt, rotate=0, color=#1, scale=.625] at (0,0) {};
\end{tikzpicture}%
}

\DeclareRobustCommand{\circle}[1]{%
\begin{tikzpicture}[ball/.style = {circle, draw, align=center, anchor=north, inner sep=0}]
   \node[ball, text width=.18cm, thick, color=#1] at (0,0) {};
\end{tikzpicture}%
}

\DeclareRobustCommand\bluecross{\cross{blue}}
\DeclareRobustCommand\purplecross{\cross{red!50!black}}
\DeclareRobustCommand\purplecircle{\circle{red!50!black}}
\DeclareRobustCommand\bluecircle{\circle{blue}}
\DeclareRobustCommand\redcircle{\circle{red}}

\begin{document}

\title{Steinmann Relations and the Wavefunction of the Universe}

\author{Paolo Benincasa}
\email[email: ]{pablowellinhouse@anche.no}
\affiliation{Niels Bohr International Academy, Niels Bohr Institute, University of Copenhagen, \\ Blegdamsvej 17, 2100 København, Denmark}
\affiliation{Instituto de F{\'i}sica Te{\'o}rica UAM/CSIC, Calle Nicol{\'a}s Cabrera 13-15,\\ Cantoblanco, 28049 Madrid, Spain}
\author{Andrew J. McLeod}
\email[email: ]{amcleod@nbi.ku.dk}
\affiliation{Niels Bohr International Academy, Niels Bohr Institute, University of Copenhagen, \\ Blegdamsvej 17, 2100 København, Denmark}
\author{Cristian Vergu}
\email[email: ]{c.vergu@gmail.com}
\affiliation{Niels Bohr International Academy, Niels Bohr Institute, University of Copenhagen, \\ Blegdamsvej 17, 2100 København, Denmark}

\begin{abstract}
The physical principles of causality and unitarity put strong constraints on the analytic structure of the flat-space S-matrix. In particular, these principles give rise to the Steinmann relations, which require that the double discontinuities of scattering amplitudes in partially-overlapping momentum channels vanish. Conversely, at cosmological scales, the imprint of causality and unitarity is in general less well understood---the wavefunction of the universe lives on the future space-like boundary, and has all time evolution integrated out. In the present work, 
we show how the flat-space Steinmann relations emerge from the structure of the wavefunction of the universe, and derive similar relations that apply to the wavefunction itself. This is done within the context of scalar toy models whose perturbative wavefunction has a first-principles definition in terms of \emph{cosmological polytopes}. In particular, we use the fact that the scattering amplitude is encoded in the \emph{scattering facet} of cosmological polytopes, and that cuts of the amplitude are encoded in the codimension-one boundaries of this facet. As we show, the flat-space Steinmann relations are thus implied by the non-existence of codimension-two boundaries at the intersection of the boundaries associated with pairs of partially-overlapping channels. 
Applying the same argument to the full cosmological polytope, we also derive Steinmann-type constraints that apply to the full wavefunction of the universe. 
These arguments show how the combinatorial properties of cosmological polytopes lead to the emergence of flat-space causality in the S-matrix, and provide new insights into the analytic structure of the wavefunction of the universe.
\end{abstract}

\maketitle

\section{Introduction}

The physics of the early universe involves processes at ultra-high energies, as the Hubble parameter during inflation can be as large as $10^{14}$ GeV. This physics is encoded in correlation functions, or equivalently in the wavefunction of the universe, living at a space-like boundary of a quasi-$dS$ spacetime at the end of inflation. Our ability to understand physics at such scales depends in part on our understanding of the analytic structure of these quantities, which we expect to be constrained by basic physical principles such as unitarity and causality. 

We can draw a parallel with what happens in flat spacetime, where the relevant quantum mechanical observable is the S-matrix. In this case there already exists a good understanding of how physical principles such as Lorentz invariance, unitarity, and causality are reflected in the structure of S-matrix elements. For instance, unitarity is encoded in the factorization properties of the S-matrix~\cite{Cutkosky:1960sp, tHooft:1973wag, Veltman:1994wz}, while causality is encoded in its analytic properties~\cite{Eden:1966dnq}.
One particular set of constraints that follow from causality and that have recently proven useful for bootstrapping amplitudes and integrals in planar $\mathcal{N}=4$ supersymmetric Yang-Mills theory~\cite{Caron-Huot:2016owq, Dixon:2016nkn,Caron-Huot:2018dsv,Drummond:2018caf,Caron-Huot:2019vjl, Caron-Huot:2019bsq,Caron-Huot:2020bkp} are the Steinmann relations~\cite{Steinmann:1960soa, Steinmann:1960sob, Araki:1961hb, Ruelle:1961rd, Stapp:1971hh, Cahill:1973px, Lassalle:1974jm, Cahill:1973qp}, which require that double discontinuities in partially overlapping channels vanish:
\begin{equation}
 \mbox{Disc}_{s_{\mathcal{I}}}\left(\mbox{Disc}_{s_{\mathcal{J}}}\mathcal{M}\right)\:=\:0 \, , \quad \text{where } \begin{cases}\mathcal{I}\nsubseteq\mathcal{J}\\ \mathcal{J}\nsubseteq\mathcal{I} \\ \mathcal{I}\cap\mathcal{J}\neq\varnothing \end{cases} \!\!\!\!. \label{eq:steinmann}
\end{equation}
Here, $\mathcal{M}$ denotes a scattering amplitude, while $\smash{s_{\mathcal{I}} = \bigl(\sum_{i \in \mathcal{I}} p_i\bigr)^2}$ is the Mandelstam invariant that depends on the external momenta whose labels belong to the set $\mathcal{I}$. The Steinmann relations apply to any quantum field theory and even to individual Feynman integrals~\cite{Bourjaily:2020wvq}---however, their implications can be subtle in cases involving massless external particles~\cite{Caron-Huot:2016owq,Bourjaily:2020wvq}.

In contrast, it is not clear how these physical principles 
make their appearance in cosmological processes, where the relevant observables are cosmological correlators or, equivalently, the wavefunctions of the universe which generate them. Such quantities live on the future space-like boundary and, consequently, have all past time evolution integrated out. So, in what ways are the imprints of causality and unitary visible?

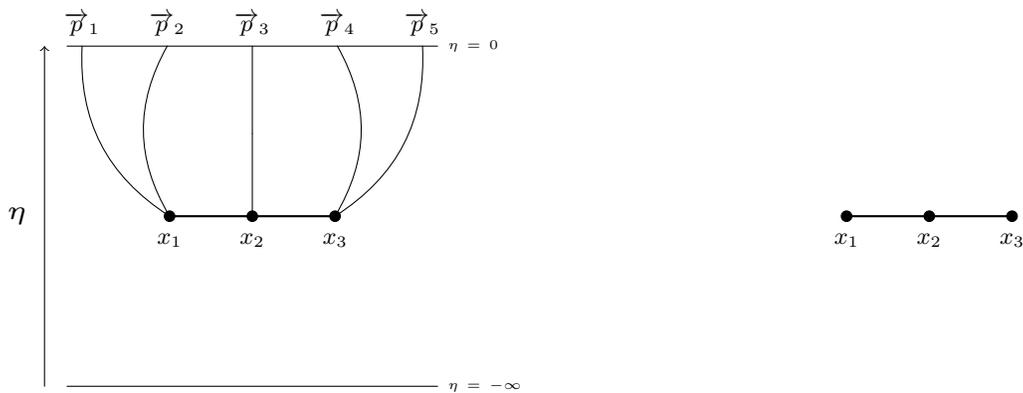
\begin{figure*}
 \centering
 \begin{tikzpicture}[line join = round, line cap = round, ball/.style = {circle, draw, align=center, anchor=north, inner sep=0}, scale=2, transform shape]
  \begin{scope}
   \def\cx{1.75}
   \def\cy{3}
   \def\r{.6}
   \pgfmathsetmacro\Axi{\cx+\r*cos(135)}
   \pgfmathsetmacro\Ayi{\cy+\r*sin(135)}
   \pgfmathsetmacro\Axf{\Axi+cos(135)}
   \pgfmathsetmacro\Ayf{\Ayi+sin(135)}
   \coordinate (pAi) at (\Axi,\Ayi);
   \coordinate (pAf) at (\Axf,\Ayf);
   \pgfmathsetmacro\Bxi{\cx+\r*cos(45)}
   \pgfmathsetmacro\Byi{\cy+\r*sin(45)}
   \pgfmathsetmacro\Bxf{\Bxi+cos(45)}
   \pgfmathsetmacro\Byf{\Byi+sin(45)}
   \coordinate (pBi) at (\Bxi,\Byi);
   \coordinate (pBf) at (\Bxf,\Byf);
   \pgfmathsetmacro\Cxi{\cx+\r*cos(-45)}
   \pgfmathsetmacro\Cyi{\cy+\r*sin(-45)}
   \pgfmathsetmacro\Cxf{\Cxi+cos(-45)}
   \pgfmathsetmacro\Cyf{\Cyi+sin(-45)}
   \coordinate (pCi) at (\Cxi,\Cyi);
   \coordinate (pCf) at (\Cxf,\Cyf);
   \pgfmathsetmacro\Dxi{\cx+\r*cos(-135)}
   \pgfmathsetmacro\Dyi{\cy+\r*sin(-135)}
   \pgfmathsetmacro\Dxf{\Dxi+cos(-135)}
   \pgfmathsetmacro\Dyf{\Dyi+sin(-135)}
   \coordinate (pDi) at (\Dxi,\Dyi);
   \coordinate (pDf) at (\Dxf,\Dyf);
   \pgfmathsetmacro\Exi{\cx+\r*cos(90)}
   \pgfmathsetmacro\Eyi{\cy+\r*sin(90)}
   \pgfmathsetmacro\Exf{\Exi+cos(90)}
   \pgfmathsetmacro\Eyf{\Eyi+sin(90)}
   \coordinate (pEi) at (\Exi,\Eyi);
   \coordinate (pEf) at (\Exf,\Eyf);
   \coordinate (ti) at ($(pDf)-(.25,0)$);
   \coordinate (tf) at ($(pAf)-(.25,0)$);
   \draw[->] (ti) -- (tf);
   \coordinate[label=left:{\tiny $\displaystyle\eta$}] (t) at ($(ti)!0.5!(tf)$);
   \coordinate (t0) at ($(pBf)+(.1,0)$);
   \coordinate (tinf) at ($(pCf)+(.1,0)$);
   \node[scale=.5, right=.0125 of t0] (t0l) {\tiny $\displaystyle\eta\,=\,0$};
   \node[scale=.5, right=.0125 of tinf] (tinfl) {\tiny $\displaystyle\eta\,=\,-\infty$};
   \draw[-] ($(pAf)-(.1,0)$) -- (t0);
   \draw[-] ($(pDf)-(.1,0)$) -- ($(pCf)+(.1,0)$);
   \coordinate (d2) at ($(pAf)!0.25!(pBf)$);
   \coordinate (d3) at ($(pAf)!0.5!(pBf)$);
   \coordinate (d4) at ($(pAf)!0.75!(pBf)$);
   \node[above=.01cm of pAf, scale=.5] (d1l) {$\displaystyle\overrightarrow{p}_1$};
   \node[above=.01cm of d2, scale=.5] (d2l) {$\displaystyle\overrightarrow{p}_2$};
   \node[above=.01cm of d3, scale=.5] (d3l) {$\displaystyle\overrightarrow{p}_3$};
   \node[above=.01cm of d4, scale=.5] (d4l) {$\displaystyle\overrightarrow{p}_4$};
   \node[above=.01cm of pBf, scale=.5] (d5l) {$\displaystyle\overrightarrow{p}_5$};
   \def\rb{.55}
   \pgfmathsetmacro\sax{\cx+\rb*cos(180)}
   \pgfmathsetmacro\say{\cy+\rb*sin(180)}
   \coordinate[label=below:{\scalebox{0.5}{$x_1$}}] (s1) at (\sax,\say);
   \pgfmathsetmacro\sbx{\cx+\rb*cos(135)}
   \pgfmathsetmacro\sby{\cy+\rb*sin(135)}
   \coordinate (s2) at (\sbx,\sby);
   \pgfmathsetmacro\scx{\cx+\rb*cos(90)}
   \pgfmathsetmacro\scy{\cy+\rb*sin(90)}
   \coordinate (s3) at (\scx,\scy);
   \pgfmathsetmacro\sdx{\cx+\rb*cos(45)}
   \pgfmathsetmacro\sdy{\cy+\rb*sin(45)}
   \coordinate (s4) at (\sdx,\sdy);
   \pgfmathsetmacro\sex{\cx+\rb*cos(0)}
   \pgfmathsetmacro\sey{\cy+\rb*sin(0)}
   \coordinate[label=below:{\scalebox{0.5}{$x_3$}}] (s5) at (\sex,\sey);
   \coordinate[label=below:{\scalebox{0.5}{$x_2$}}] (sc) at (\cx,\cy);
   \draw (s1) edge [bend left] (pAf);
   \draw (s1) edge [bend left] (d2);
   \draw (s3) -- (d3);
   \draw (s5) edge [bend right] (d4);
   \draw (s5) edge [bend right] (pBf);
   \draw [fill] (s1) circle (1pt);
   \draw [fill] (sc) circle (1pt);
   \draw (s3) -- (sc);
   \draw [fill] (s5) circle (1pt);
   \draw[-,thick] (s1) -- (sc) -- (s5);
  \end{scope}
  \begin{scope}[shift={(4.5,0)}, transform shape]
   \def\cx{1.75}
   \def\cy{3}
   \def\r{.6}
   \pgfmathsetmacro\Axi{\cx+\r*cos(135)}
   \pgfmathsetmacro\Ayi{\cy+\r*sin(135)}
   \pgfmathsetmacro\Axf{\Axi+cos(135)}
   \pgfmathsetmacro\Ayf{\Ayi+sin(135)}
   \coordinate (pAi) at (\Axi,\Ayi);
   \coordinate (pAf) at (\Axf,\Ayf);
   \pgfmathsetmacro\Bxi{\cx+\r*cos(45)}
   \pgfmathsetmacro\Byi{\cy+\r*sin(45)}
   \pgfmathsetmacro\Bxf{\Bxi+cos(45)}
   \pgfmathsetmacro\Byf{\Byi+sin(45)}
   \coordinate (pBi) at (\Bxi,\Byi);
   \coordinate (pBf) at (\Bxf,\Byf);
   \pgfmathsetmacro\Cxi{\cx+\r*cos(-45)}
   \pgfmathsetmacro\Cyi{\cy+\r*sin(-45)}
   \pgfmathsetmacro\Cxf{\Cxi+cos(-45)}
   \pgfmathsetmacro\Cyf{\Cyi+sin(-45)}
   \coordinate (pCi) at (\Cxi,\Cyi);
   \coordinate (pCf) at (\Cxf,\Cyf);
   \pgfmathsetmacro\Dxi{\cx+\r*cos(-135)}
   \pgfmathsetmacro\Dyi{\cy+\r*sin(-135)}
   \pgfmathsetmacro\Dxf{\Dxi+cos(-135)}
   \pgfmathsetmacro\Dyf{\Dyi+sin(-135)}
   \coordinate (pDi) at (\Dxi,\Dyi);
   \coordinate (pDf) at (\Dxf,\Dyf);
   \pgfmathsetmacro\Exi{\cx+\r*cos(90)}
   \pgfmathsetmacro\Eyi{\cy+\r*sin(90)}
   \pgfmathsetmacro\Exf{\Exi+cos(90)}
   \pgfmathsetmacro\Eyf{\Eyi+sin(90)}
   \coordinate (pEi) at (\Exi,\Eyi);
   \coordinate (pEf) at (\Exf,\Eyf);
   \coordinate (ti) at ($(pDf)-(.25,0)$);
   \coordinate (tf) at ($(pAf)-(.25,0)$);
   \def\rb{.55}
   \pgfmathsetmacro\sax{\cx+\rb*cos(180)}
   \pgfmathsetmacro\say{\cy+\rb*sin(180)}
   \coordinate[label=below:{\scalebox{0.5}{$x_1$}}] (s1) at (\sax,\say);
   \pgfmathsetmacro\sbx{\cx+\rb*cos(135)}
   \pgfmathsetmacro\sby{\cy+\rb*sin(135)}
   \coordinate (s2) at (\sbx,\sby);
   \pgfmathsetmacro\scx{\cx+\rb*cos(90)}
   \pgfmathsetmacro\scy{\cy+\rb*sin(90)}
   \coordinate (s3) at (\scx,\scy);
   \pgfmathsetmacro\sdx{\cx+\rb*cos(45)}
   \pgfmathsetmacro\sdy{\cy+\rb*sin(45)}
   \coordinate (s4) at (\sdx,\sdy);
   \pgfmathsetmacro\sex{\cx+\rb*cos(0)}
   \pgfmathsetmacro\sey{\cy+\rb*sin(0)}
   \coordinate[label=below:{\scalebox{0.5}{$x_3$}}] (s5) at (\sex,\sey);
   \coordinate[label=below:{\scalebox{0.5}{$x_2$}}] (sc) at (\cx,\cy);
   \draw [fill] (s1) circle (1pt);
   \draw [fill] (sc) circle (1pt);
   \draw [fill] (s5) circle (1pt);
   \draw[-,thick] (s1) -- (sc) -- (s5);
  \end{scope}
 \end{tikzpicture}
 \caption{On the left, we show a Feynman graph that contributes to the wavefunction of the universe. On the right, we depict the associated reduced graph, which is obtained from the Feynman graph by suppressing the external edges.}
 \label{fig:G}
\end{figure*}

A similar question---regarding how flat-space unitarity and Lorentz invariance emerge from the wavefunction of the universe---was discussed in \cite{Arkani-Hamed:2018bjr} in the context of
\emph{cosmological polytopes}~\cite{Arkani-Hamed:2017fdk, Benincasa:2019vqr}. Cosmological polytopes are geometrical-combinatorial objects 
which encode all of the properties of the wavefunction despite having a first-principles definition that makes no reference to physical concepts such as Hilbert space or spacetime. In particular, they constitute a special class of positive geometries, which come equipped with a {\it canonical form} that has logarithmic singularities (only) along the boundaries of the polytope.\footnote{We emphasize that positive geometries can be defined and studied independently of any physical interpretation \cite{Arkani-Hamed:2017tmz}.}  In the case of cosmological polytopes, these singularities come in one-to-one correspondence with the singularities of the wavefunction, and each of them gives rise to an associated {\it canonical function} that provides the contribution of
a Feynman graph $\mathcal{G}$ to the wavefunction itself~\cite{Arkani-Hamed:2017fdk}.
One of these singularities occurs at the vanishing locus of the total energy $E_{\mbox{\tiny tot}} = \sum_i |\vec{p}_i|$, where $\vec{p}_i$ is the spatial-momentum of the $i^{\text{th}}$ external state (note that for physical processes 
$|\vec{p}_i|$ is always positive and, hence, such locus can be reached upon analytic continuation only). As such a singularity is approached, the wavefunction of the universe reduces to (the high energy limit of) the flat-space scattering amplitude\footnote{When the scattering amplitude vanishes or the states under consideration do not have a flat-space counterpart, the total energy singularity is softer and its coefficient is a typical cosmological effect---see \cite{Grall:2020ibl} for an example.} \cite{Maldacena:2011nz, Raju:2012zr,Arkani-Hamed:2015bza}. In the cosmological polytope picture, the total energy singularity identifies the {\it scattering facet}, which is a polytope living on a codimension-one boundary of the cosmological polytope, whose canonical form returns the relevant scattering amplitude \cite{Arkani-Hamed:2017fdk}. The vertex structure of this facet makes flat-space unitarity manifest;
in particular, the codimension-one faces of the scattering facet themselves factorize into pairs of lower-dimensional scattering facets and a simplex that encodes the Lorentz-invariant phase-space measure. This provides a combinatorial statement of the cutting rules, which encode unitarity. In a similar manner, Lorentz invariance on the scattering facet is made manifest by a contour integral representation of the canonical form~\cite{Arkani-Hamed:2018bjr}.

In this work, we adopt a similar strategy for studying how the flat-space causality is encoded in the wavefunction of the universe. In particular, we investigate how the Steinmann relations are encoded in the scattering facet, utilizing the correspondence between the boundaries of these polytopes and the cuts of Feynman integrals.\footnote{The connection between individual discontinuities and cut integrals has long been understood~\cite{cutkosky}. For more details on how sequential discontinuities of scattering amplitudes can be computed using cut integrals, see~\cite{Abreu:2014cla,Bourjaily:2020wvq} (also~\cite{Bloch:2015efx,Abreu:2017ptx}).} 
We also ask whether Steinmann-like relations hold for the wavefunction itself, whose cuts now correspond to facets of the full cosmological polytope.

As we review below, for both the scattering amplitude and the wavefunction of the universe, these cuts are more specifically encoded in the residues of the canonical form $\omega(\mathcal{P})$ on the boundaries of the associated polytope $\mathcal{P}$, where $\mathcal{P}$ is either the full cosmological polytope $\mathcal{P}_{\mathcal{G}}$ or the scattering facet $\mathcal{S}_{\mathcal{G}}$. These boundaries come in one-to-one correspondence with the subgraphs $\mathfrak{g} \subset \mathcal{G}$, and can be characterized by hyperplanes \(\mathcal{W}_{\mathfrak{g}}\) in dual projective space; thus, we denote the residue of $\omega(\mathcal{P})$ on the boundary corresponding to $\mathfrak{g}$ by $\operatorname{Res}_{\mathcal{W}_{\mathfrak{g}}} \omega(\mathcal{P})$.

The canonical form has the property that its residue along the hypersurface $\mathcal{W}_{\mathfrak{g}}$ itself constitutes the canonical form on this boundary of $\mathcal{P}$:
\begin{equation}
    \omega(\mathcal{P} \cap \mathcal{W}_{\mathfrak{g}}) = \operatorname{Res}_{\mathcal{W}_{\mathfrak{g}}} \omega(\mathcal{P}).
\end{equation}
Importantly, this new canonical form only has singularities along the hyperplanes in \(\mathcal{W}_{\mathfrak{g}}\) which contain its facets.  In other words,
\begin{equation} \label{eq:double_residue_intro}
    \operatorname{Res}_{\mathcal{W}_{\mathfrak{g}_1}} \operatorname{Res}_{\mathcal{W}_{\mathfrak{g}_2}} \omega(\mathcal{P}) = 0
\end{equation}
if \(\mathcal{W}_{\mathfrak{g}_1} \cap \mathcal{W}_{\mathfrak{g}_2}\) does not contain a codimension-two facet of \(\mathcal{P}\).  Below, we will show that the resemblance between equations~\eqref{eq:steinmann} and~\eqref{eq:double_residue_intro} is not a coincidence.

In what follows, we first review the wavefunction of the universe and the salient aspects of cosmological polytopes. We then investigate the question of how flat-space causality emerges on the scattering facet of cosmological polytopes, where the Steinmann relations are known to apply. We demonstrate that the Steinmann relations emerge naturally from the face structure of
the scattering facet, insofar as the codimension-one boundaries  of this facet that correspond to partially-overlapping momentum channels never intersect to form codimension-two boundaries. This elucidates the mechanism by which flat-space causality emerges from the wavefunction of the universe, complementing the existing understanding of how flat-space  unitarity and Lorentz invariance also emerge~\cite{Arkani-Hamed:2018bjr}. Having uncovered the combinatorial mechanism by which these relations are encoded, we go on to derive novel constraints on the wavefunction of the universe that follow from the same argument. 


\section{Cosmological polytopes} \label{sec:WFCP}

Let us consider a massless scalar theory in $(d+1)$-dimensional flat spacetime with polynomial interactions that have time-dependent couplings,
\begin{equation}
 S[\phi]\:=\:\int d^d x\,\int_{-\infty}^0d\eta\,
  \left[
   \frac{1}{2}\left(\partial\phi\right)^2-\sum_{k\ge3}\frac{\lambda_k(\eta)}{k!}\phi^k
  \right].
\end{equation}
This theory describes a conformally-coupled scalar in an FRW cosmology, where $d s^2\,=\,a^2(\eta)[-d\eta^2+\delta_{ij}dx^id x^j]$, provided that the couplings $\lambda_k(\eta)$ are taken to be 
$$\lambda_k(\eta)\,=\,\lambda_k\vartheta(-\eta)[a(\eta)]^{(2-k)(d-1)/2+2}$$
for some constants $\lambda_k$.

The wavefunction of the universe for this theory is given by
\begin{equation}
 \Psi[\Phi]=\hspace{-.25cm}\int\limits_{\phi(-\infty(1-i\varepsilon))\,=\,0}^{\phi(0) = \Phi} \hspace{-.25cm}D\phi\:e^{i S[\phi]},
\end{equation}
where the $i\varepsilon$ prescription regularizes the path integral at early times, and $\Phi$ corresponds to the boundary state at late time. We split $\phi$ into a classical mode and a quantum fluctuation via  
\begin{equation}
\phi(\vec{p},\eta) = \Phi(\vec{p}) e^{i |\vec{p}| \eta} \,+\, \varphi \, ,
\end{equation}
where the classical part of the solution has the correct Bunch-Davies oscillatory behavior at early times, and we require the fluctuations $\varphi$ to vanish at both early and late times ($\eta=-\infty,\,0$ respectively). 

After Fourier-transforming the couplings as
\begin{equation}
    \lambda_k(\eta)=\int_{-\infty}^{+\infty}d\epsilon\,e^{i\epsilon\eta}\tilde{\lambda}_k(\epsilon) \, , \label{eq:fourier_couplings}
\end{equation}
the perturbative wavefunction can be computed in terms of Feynman diagrams, which now include an integral over the energy $\epsilon$ associated with each graph site.\footnote{In order to avoid language clashes, we will refer to the vertices of the graphs as {\it sites}, reserving the name {\it vertices} for the polytopes.} We here postpone consideration of these integrals over site energies,\footnote{For recent work on the integrals over site energies, see~\cite{Hillman:2019wgh}.} as well as the integrals over the spatial
loop momenta in the case of loop graphs,
and focus on the remaining contribution coming from a graph $\mathcal{G}$ with sites $\mathcal{V}$ and edges $\mathcal{E}$,
\begin{equation}\label{eq:WFint}
 \!\!\! \psi_{\mathcal{G}}(x_v,y_e)=\! \int_{-\infty}^0\prod_{v\in\mathcal{V}}\left[d\eta_v\,e^{ix_v\eta_v}\right]\prod_{e\in\mathcal{E}}G(\eta_{v_e},\eta_{v'_e},y_e),
\end{equation}
where $x_v = \sum_{i \in v} |\vec{p}_i|$ is the sum of external energies entering site $v \in \mathcal{V}$, $y_e$ is the energy flowing through edge $e \in \mathcal{E}$, and 
\begin{align}
G(\eta_{v_e},\eta_{v'_e},y_e)=(2y_e)^{-1}&\big[e^{-i y_e(\eta_{v_e}-\eta_{v'_e})}\vartheta(\eta_{v_e}-\eta_{v'_e}) \nonumber \\ 
&\quad +e^{+i y_e(\eta_{v_e}-\eta_{v_e'})}\vartheta(\eta_{v'_e}-\eta_{v_e})\nonumber \\
&\quad -e^{i y_e(\eta_{v_e}+\eta_{v'_e})}\big] 
\end{align}
is the bulk-to-bulk propagator that satisfies the boundary condition that all fluctuations vanish at $\eta=0$.\footnote{While the energy flowing through each edge will be partially fixed by momentum conservation, we leave these energies general as the function $\psi_{\mathcal{G}}(x_v, y_e)$ doesn't know about these constraints.} 
This contribution to the Feynman diagram is universal, insofar as the details of the specific theory are entirely encoded in the Fourier coefficients $\tilde{\lambda}_k(\epsilon)$, which we have factored out as part of the energy integrals. For further details see \cite{Arkani-Hamed:2017fdk}. 

\begin{figure*}[t]
 \begin{tikzpicture}[line join = round, line cap = round, ball/.style = {circle, draw, align=center, anchor=north, inner sep=0},
                     axis/.style={very thick, ->, >=stealth'}, pile/.style={thick, ->, >=stealth', shorten <=2pt, shorten>=2pt}, every node/.style={color=black}]
  \begin{scope}[scale={.5}, shift={(6,3)}, transform shape]
   \pgfmathsetmacro{\factor}{1/sqrt(2)};
   \coordinate  (B2) at (1.5,-3,-1.5*\factor);
   \coordinate  (A1) at (-1.5,-3,-1.5*\factor);
   \coordinate  (B1) at (1.5,-3.75,1.5*\factor);
   \coordinate  (A2) at (-1.5,-3.75,1.5*\factor);
   \coordinate  (C1) at (0.75,-.65,.75*\factor);
   \coordinate  (C2) at (0.4,-6.05,.75*\factor);
   \coordinate (Int) at (intersection of A2--B2 and B1--C1);
   \coordinate (Int2) at (intersection of A1--B1 and A2--B2);

   \tikzstyle{interrupt}=[
    postaction={
        decorate,
        decoration={markings,
                    mark= at position 0.5
                          with
                          {
                            \node[rectangle, color=white, fill=white, below=-.1 of Int] {};
                          }}}
   ]

   \draw[interrupt,thick,color=red] (B1) -- (C1);
   \draw[-,very thick,color=blue] (A1) -- (B1);
   \draw[-,very thick,color=blue] (A2) -- (B2);
   \draw[-,very thick,color=blue] (A1) -- (C1);
   \draw[-, dashed, very thick, color=red] (A2) -- (C2);
   \draw[-, dashed, thick, color=blue] (B2) -- (C2);

   \coordinate[label=below:{\Large ${\bf x'}_i$}] (x2) at ($(A1)!0.5!(B1)$);
   \draw[fill,color=blue] (x2) circle (2.5pt);
   \coordinate[label=left:{\Large ${\bf x}_i$}] (x1) at ($(C1)!0.5!(A1)$);
   \draw[fill,color=blue] (x1) circle (2.5pt);
   \coordinate[label=right:{\Large ${\bf x}_j$}] (x3) at ($(B2)!0.5!(C2)$);
   \draw[fill,color=blue] (x3) circle (2.5pt);
  \end{scope}
  \begin{scope}[scale={.5}, shift={(15,3)}, transform shape]
   \pgfmathsetmacro{\factor}{1/sqrt(2)};  
   \coordinate (B2c) at (1.5,-3,-1.5*\factor);
   \coordinate (A1c) at (-1.5,-3,-1.5*\factor);
   \coordinate (B1c) at (1.5,-3.75,1.5*\factor);
   \coordinate (A2c) at (-1.5,-3.75,1.5*\factor);  
   \coordinate (C1c) at (0.75,-.65,.75*\factor);
   \coordinate (C2c) at (0.4,-6.05,.75*\factor);
   \coordinate (Int3) at (intersection of A2c--B2c and B1c--C1c);

   \node at (A1c) (A1d) {};
   \node at (B2c) (B2d) {};
   \node at (B1c) (B1d) {};
   \node at (A2c) (A2d) {};
   \node at (C1c) (C1d) {};
   \node at (C2c) (C2d) {};

   \draw[-,dashed,fill=blue!30, opacity=.7] (A1c) -- (B2c) -- (C1c) -- cycle;
   \draw[-,thick,fill=blue!20, opacity=.7] (A1c) -- (A2c) -- (C1c) -- cycle;
   \draw[-,thick,fill=blue!20, opacity=.7] (B1c) -- (B2c) -- (C1c) -- cycle;
   \draw[-,thick,fill=blue!35, opacity=.7] (A2c) -- (B1c) -- (C1c) -- cycle;

   \draw[-,dashed,fill=red!30, opacity=.3] (A1c) -- (B2c) -- (C2c) -- cycle;
   \draw[-,dashed, thick, fill=red!50, opacity=.5] (B2c) -- (B1c) -- (C2c) -- cycle;
   \draw[-,dashed,fill=red!40, opacity=.3] (A1c) -- (A2c) -- (C2c) -- cycle;
   \draw[-,dashed, thick, fill=red!45, opacity=.5] (A2c) -- (B1c) -- (C2c) -- cycle;
  \end{scope}
  \begin{scope}[scale={.6}, shift={(5,-3.5)}, transform shape]
   \pgfmathsetmacro{\factor}{1/sqrt(2)};
   \coordinate  (c1b) at (0.75,0,-.75*\factor);
   \coordinate  (b1b) at (-.75,0,-.75*\factor);
   \coordinate  (a2b) at (0.75,-.65,.75*\factor);

   \coordinate  (c2b) at (1.5,-3,-1.5*\factor);
   \coordinate  (b2b) at (-1.5,-3,-1.5*\factor);
   \coordinate  (a1b) at (1.5,-3.75,1.5*\factor);

   \coordinate (Int1) at (intersection of b2b--c2b and b1b--a1b);
   \coordinate (Int2) at (intersection of b2b--c2b and c1b--a1b);
   \coordinate (Int3) at (intersection of b2b--a2b and b1b--a1b);
   \coordinate (Int4) at (intersection of a2b--c2b and c1b--a1b);
   \tikzstyle{interrupt}=[
    postaction={
        decorate,
        decoration={markings,
                    mark= at position 0.5
                          with
                          {
                            \node[rectangle, color=white, fill=white] at (Int1) {};
                            \node[rectangle, color=white, fill=white] at (Int2) {};
                          }}}
   ]

   \node at (c1b) (c1c) {};
   \node at (b1b) (b1c) {};
   \node at (a2b) (a2c) {};
   \node at (c2b) (c2c) {};
   \node at (b2b) (b2c) {};
   \node at (a1b) (a1c) {};

   \draw[interrupt,thick,color=red] (b2b) -- (c2b);
   \draw[-,very thick,color=red] (b1b) -- (c1b);
   \draw[-,very thick,color=blue] (b1b) -- (a1b);
   \draw[-,very thick,color=blue] (a1b) -- (c1b);
   \draw[-,very thick,color=blue] (b2b) -- (a2b);
   \draw[-,very thick,color=blue] (a2b) -- (c2b);

   \node[ball,text width=.15cm,fill,color=blue, above=-.06cm of Int3, label=left:{\large ${\bf x}_i$}] (Inta) {};
   \node[ball,text width=.15cm,fill,color=blue, above=-.06cm of Int4, label=right:{\large ${\bf x'}_i$}] (Intb) {};

  \end{scope}
  \begin{scope}[scale={.6}, shift={(12.5,-3.5)}, transform shape]
   \pgfmathsetmacro{\factor}{1/sqrt(2)};
   \coordinate (c1a) at (0.75,0,-.75*\factor);
   \coordinate (b1a) at (-.75,0,-.75*\factor);
   \coordinate (a2a) at (0.75,-.65,.75*\factor);
  
   \coordinate (c2a) at (1.5,-3,-1.5*\factor);
   \coordinate (b2a) at (-1.5,-3,-1.5*\factor);
   \coordinate (a1a) at (1.5,-3.75,1.5*\factor);

   \draw[-,dashed,fill=green!50,opacity=.6] (c1a) -- (b1a) -- (b2a) -- (c2a) -- cycle;
   \draw[draw=none,fill=red!60, opacity=.45] (c2a) -- (b2a) -- (a1a) -- cycle;
   \draw[-,fill=blue!,opacity=.3] (c1a) -- (b1a) -- (a2a) -- cycle; 
   \draw[-,fill=green!50,opacity=.4] (b1a) -- (a2a) -- (a1a) -- (b2a) -- cycle;
   \draw[-,fill=green!45!black,opacity=.2] (c1a) -- (a2a) -- (a1a) -- (c2a) -- cycle;  
  \end{scope}
  \begin{scope}[shift={(11,0)}, transform shape]
   \coordinate[label=below:{\footnotesize $x_i$}] (x1) at (0,0);
   \coordinate[label=below:{\footnotesize $x'_i$}] (x2) at ($(x1)+(1.5,0)$);
   \coordinate[label=above:{\footnotesize $y_i$}] (yi) at ($(x1)!0.5!(x2)$);
   \coordinate[label=below:{\footnotesize $x_j$}] (x3) at ($(x2)+(1.5,0)$);
   \coordinate[label=above:{\footnotesize $y_j$}] (yj) at ($(x2)!0.5!(x3)$);

   \draw[-, very thick, color=red] (x1) -- (x2) -- (x3);
   \draw[color=blue,fill=blue] (x1) circle (2pt);
   \draw[color=blue,fill=blue] (x2) circle (2pt);
   \draw[color=blue,fill=blue] (x3) circle (2pt);
  \end{scope}
  \begin{scope}[shift={(11,-3)}, transform shape]
   \coordinate[label=left:{\footnotesize $x_i$}] (x1) at (.75,0);
   \coordinate[label=right:{\footnotesize $x_j$}] (x2) at ($(x1)+(1.5,0)$);
   \coordinate (c) at ($(x1)!0.5!(x2)$);
   \coordinate[label=above:{\footnotesize $y_i$}] (yi) at ($(c)+(0,.75)$);
   \coordinate[label=below:{\footnotesize $y_j$}] (yj) at ($(c)-(0,.75)$);
   
   \draw[color=red, very thick] (c) circle (.75cm);
   \draw[color=blue,fill=blue] (x1) circle (2pt);
   \draw[color=blue,fill=blue] (x2) circle (2pt);
  \end{scope}
 \end{tikzpicture}
\caption{Examples of cosmological polytopes obtained from the intersection of two triangles. The images in the left column illustrate the intersection of pairs of triangles at either one or two midpoints, while the corresponding convex hulls are depicted in the middle column. The column on the right shows the associated reduced graphs.}
 \label{fig:cp}
\end{figure*}
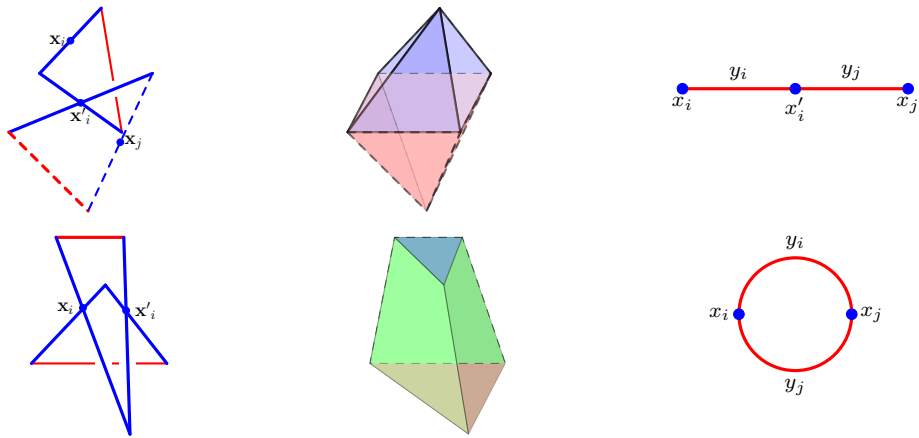

Since the integrals over conformal time in equation~\eqref{eq:WFint} only depend on the total energy entering each site, they can be represented by reduced graphs in which the lines connected to external states have been suppressed. An example Feynman graph and the associated reduced graph are shown in Figure~\ref{fig:G}. In~\cite{Arkani-Hamed:2017fdk} it was shown that these reduced graphs are in one-to-one correspondence with polytopes whose canonical forms encode the same singularity structure as $\psi_{\mathcal{G}}(x_v,y_e)$. To identify the polytope associated with a given graph $\mathcal{G}$, we associate a triangle with each of its propagators, where the edges of this triangle can be thought of as representing the two sites and the edge that constitute this propagator:
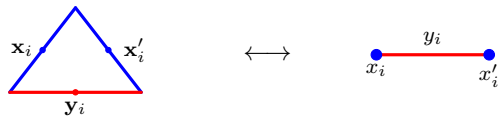
\begin{equation*}
  \begin{tikzpicture}[line join = round, line cap = round, ball/.style = {circle, draw, align=center, anchor=north, inner sep=0},
                     axis/.style={very thick, ->, >=stealth'}, pile/.style={thick, ->, >=stealth', shorten <=2pt, shorten>=2pt}, every node/.style={color=black}]
  \begin{scope}[scale={.5}]
   \coordinate (A) at (0,0);
   \coordinate (B) at (-1.75,-2.25);
   \coordinate (C) at (+1.75,-2.25);
   \coordinate [label=left:{\footnotesize $\displaystyle {\bf x}_i$}] (m1) at ($(A)!0.5!(B)$);
   \coordinate [label=right:{\footnotesize $\displaystyle \;{\bf x}'_i$}] (m2) at ($(A)!0.5!(C)$);
   \coordinate [label=below:{\footnotesize $\displaystyle {\bf y}_i$}] (m3) at ($(B)!0.5!(C)$);
   \tikzset{point/.style={insert path={ node[scale=2.5*sqrt(\pgflinewidth)]{.} }}}

   \draw[color=blue,fill=blue] (m1) circle (2pt);
   \draw[color=blue,fill=blue] (m2) circle (2pt);
   \draw[color=red,fill=red] (m3) circle (2pt);

   \draw[-, very thick, color=blue] (B) -- (A);
   \draw[-, very thick, color=blue] (A) -- (C);
   \draw[-, very thick, color=red] (B) -- (C);
  \end{scope}
  \begin{scope}
   \coordinate[label=below:{\footnotesize $x_i$}] (x1) at (4,-.625);
   \coordinate[label=below:{\footnotesize $x'_i$}] (x2) at ($(x1)+(1.5,0)$);
   \coordinate[label=above:{\footnotesize $y_i$}] (y) at ($(x1)!0.5!(x2)$);

   \draw[-, very thick, color=red] (x1) -- (x2);
   \draw[color=blue,fill=blue] (x1) circle (2pt);
   \draw[color=blue,fill=blue] (x2) circle (2pt);

   \node[left=1cm of x1] {$\displaystyle\longleftrightarrow$};
  \end{scope}
 \end{tikzpicture}
\end{equation*}
We can situate these triangles in projective space by taking $\mathbf{x}_i$, $\mathbf{y}_i$, and $\mathbf{x}'_i$ to be generic vectors in $\mathbb{P}^{3n_e-1}$, where $n_e$ is the number of propagators in $\mathcal{G}$. The vertices of the $i^{\text{th}}$ triangle will then be given by the vectors $\mathbf{x}_i-\mathbf{y}_i+\mathbf{x}'_i$, $\mathbf{x}_i+\mathbf{y}_i-\mathbf{x}'_i$, and $-\mathbf{x}_i+\mathbf{y}_i+\mathbf{x}'_i$. However, when multiple propagators end on the same graph site, the corresponding midpoints of these triangles should be identified, which implies that these triangles will be contained within a lower-dimensional subspace of $\mathbb{P}^{3n_e-1}$. In particular, the vertices appearing in this collection of triangles will only span a subspace of dimension $n_v+n_e-1$, where $n_v$ and $n_e$ represent the number of sites and edges in $\mathcal{G}$. The cosmological polytope $\mathcal{P}_{\mathcal{G}}$ associated with $\mathcal{G}$ is defined to be the convex hull of these triangle vertices. Examples are shown in Figure~\ref{fig:cp}.

Let us illustrate the construction described above on the simple case of a triangle \(\Delta_i\), whose edge midpoints are at $\mathbf{x}_i$, $\mathbf{y}_i$, and $\mathbf{x'}_i$.  We can embed the triangle in the projective plane via the identifications \(\mathbf{x}_i = [1 : 0 : 0]\), \(\mathbf{y}_i = [0 : 1 : 0]\) and \(\mathbf{x'}_i = [0 : 0 : 1]\).  Then, we can parametrize a generic point within the corresponding cosmological polytope by $\mathcal{Y} = x_i \mathbf{x}_i + y_i \mathbf{y}_i + x_i' \mathbf{x'}_i \in \Delta_i$, where $x_i$, $y_i$, and $x'_i$ are real numbers that satisfy \(x_i + y_i \geq 0\), \(x_i + x_i' \geq 0\) and \(y_i + x_i' \geq 0\).

More generally, given a vector $\mathcal{Y}$ inside a cosmological polytope \(\mathcal{P_G}\), we can formulate a unique canonical form \begin{equation}
  \omega(\mathcal{Y},\mathcal{P}_{\mathcal{G}})\,=\,\Omega(\mathcal{Y},\mathcal{P}_{\mathcal{G}})\langle\mathcal{Y}d^{N}\mathcal{Y}\rangle  
\end{equation} in $\mathcal{P}_{\mathcal{G}}$.\footnote{Note that we sometimes leave the dependence of the canonical form and the canonical function on $\mathcal{Y}$ implicit.} This form has logarithmic singularities on---and only on---all of its boundaries~\cite{Arkani-Hamed:2017fdk}. Moreover, $\Omega(\mathcal{Y},\mathcal{P}_{\mathcal{G}})$ is called the associated {\it canonical function}, and turns out to be precisely the universal part of the Feynman diagram from equation~\eqref{eq:WFint},\footnote{In cases involving internal massless states in a generic FRW cosmology, the function $\psi_{\mathcal{G}}(x_v,\,y_e)$ is instead determined by a {\it covariant form} associated to the very same cosmological polytope~\cite{Benincasa:2019vqr, Benincasa:2020uph}.} namely
\begin{equation}\label{eq:cfwf}
 \Omega(\mathcal{Y},\,\mathcal{P}_{\mathcal{G}})\:=\:\psi_{\mathcal{G}}(x_v,\,y_e).
\end{equation}
One of the general features of so-called canonical functions is that they are singular (only) on the boundaries of the associated positive geometry. Thus, by virtue of equation~\eqref{eq:cfwf}, the boundary structure of $\mathcal{P}_{\mathcal{G}}$ characterises the residues of $\psi_{\mathcal{G}}(x_v,\,y_e)$. Moreover, as shown in~\cite{Arkani-Hamed:2017fdk}, the codimension-one boundaries (or facets) of $\mathcal{P}_{\mathcal{G}}$ are in one-to-one correspondence with the connected subgraphs of \(\mathcal{G}\). In particular, each of the codimension-one facets of $\mathcal{P}_{\mathcal{G}}$ can be characterized by a dual vector 
\begin{equation} \label{eq:dual_vec}
        \mathcal{W}_{\mathfrak{g}} = \sum_{v \in \mathfrak{g}} \tilde{\mathbf{x}}_v + \sum_{e \in \mathcal{E}_{\mathfrak{g}}^{\text{ext}}} \tilde{\mathbf{y}}_e\, ,
\end{equation}
where $\mathfrak{g}$ is one of the connected subgraphs of $\mathcal{G}$, \(\tilde{\mathbf{x}}_v\) and \(\tilde{\mathbf{y}}_e\) are dual vectors to \(\mathbf{x}_v\) and \(\mathbf{y}_e\) (meaning \(\tilde{\mathbf{x}}_v \cdot \mathbf{x}_{v'} = \delta_{v v'}\), \(\tilde{\mathbf{y}}_e \cdot \mathbf{y}_{e'} = \mathbf{\delta}_{e e'}\), and \(\tilde{\mathbf{x}}_v \cdot \mathbf{y}_e = \tilde{\mathbf{y}}_e \cdot \mathbf{x}_v = 0\)), and \(\mathcal{E}_{\mathfrak{g}}^{\text{ext}}\) is the set of edges departing from the subgraph \(\mathfrak{g}\). It can be checked that \(\mathcal{W}_{\mathfrak{g}} \cdot \mathcal{Y} \geq 0\) for all points \(\mathcal{Y} \in \mathcal{P_G}\). We also associate an energy to the subgraph $\mathfrak{g}$, namely
\begin{equation} \label{eq:subraph_energy}
    E_{\mathfrak{g}} = \sum_{v \in \mathfrak{g}} x_v + \sum_{e \in \mathcal{E}_{\mathfrak{g}}^{\text{ext}}} y_e \, ,
\end{equation}
and we have that $E_{\mathfrak{g}} \to 0$ as we approach the corresponding facet.

The correspondence between polytopes and graphs allows us to keep track of the vertex structure of any facet by introducing a simple marking which identifies those vertices which {\it are not} on the facet:
\begin{equation*}
 \begin{tikzpicture}[ball/.style = {circle, draw, align=center, anchor=north, inner sep=0}, cross/.style={cross out, draw, minimum size=2*(#1-\pgflinewidth), inner sep=0pt, outer sep=0pt}]
  \begin{scope}
   \coordinate[label=below:{\footnotesize $x_i$}] (x1) at (0,0);
   \coordinate[label=below:{\footnotesize $x_i'$}] (x2) at ($(x1)+(1.5,0)$);
   \coordinate[label=below:{\footnotesize $y_{e}$}] (y) at ($(x1)!0.5!(x2)$);
   \draw[-,very thick] (x1) -- (x2);
   \draw[fill=black] (x1) circle (2pt);
   \draw[fill=black] (x2) circle (2pt);
   \node[very thick, cross=4pt, rotate=0, color=blue] at (y) {};   
   \node[below=.375cm of y, scale=.7] (lb1) {$\mathcal{W}\cdot({\bf x}_i-{\bf y}_{e}+{\bf x}'_{i})>\,0$};  
  \end{scope}
  \begin{scope}[shift={(3,0)}]
   \coordinate[label=below:{\footnotesize $x_i$}] (x1) at (0,0);
   \coordinate[label=below:{\footnotesize $x_i'$}] (x2) at ($(x1)+(1.5,0)$);
   \coordinate[label=below:{\footnotesize $y_{e}$}] (y) at ($(x1)!0.5!(x2)$);
   \draw[-,very thick] (x1) -- (x2);
   \draw[fill=black] (x1) circle (2pt);
   \draw[fill=black] (x2) circle (2pt);
   \node[very thick, cross=4pt, rotate=0, color=blue] at ($(x1)+(.25,0)$) {};   
   \node[below=.375cm of y, scale=.7] (lb1) {$\mathcal{W}\cdot({\bf x}_i+{\bf y}_{e}-{\bf x}'_{i})>\,0$};  
  \end{scope}
  \begin{scope}[shift={(6,0)}]
   \coordinate[label=below:{\footnotesize $x_i$}] (x1) at (0,0);
   \coordinate[label=below:{\footnotesize $x_i'$}] (x2) at ($(x1)+(1.5,0)$);
   \coordinate[label=below:{\footnotesize $y_{e}$}] (y) at ($(x1)!0.5!(x2)$);
   \draw[-,very thick] (x1) -- (x2);
   \draw[fill=black] (x1) circle (2pt);
   \draw[fill=black] (x2) circle (2pt);
   \node[very thick, cross=4pt, rotate=0, color=blue] at ($(x2)-(.25,0)$) {};   
   \node[below=.375cm of y, scale=.7] (lb1) {$\mathcal{W}\cdot(-{\bf x}_i+{\bf y}_{e}+{\bf x}'_{i})>\,0$};  
  \end{scope}
 \end{tikzpicture}
\end{equation*}
In particular, the facet of $\mathcal{P}_{\mathcal{G}}$ identified with a subgraph $\mathfrak{g}\subseteq\mathcal{G}$ is given by marking all internal edges of $\mathfrak{g}$ in the middle, and all the edges that depart this subgraph on the side closest to $\mathfrak{g}$. This association fully characterises the facets of $\mathcal{P}_{\mathcal{G}}$.

Finally, let us highlight that the combinatorial structure of cosmological polytopes
is such that each of their codimension-$k$ faces is given by the intersection of a pair of codimension-$(k{-}1)$ faces. Given that $\mathcal{P}_{\mathcal{G}}$ has dimension $n_v+n_e-1$, the intersection of two of these codimension-$(k{-}1)$ faces can only have codimension $k$ if the intersecting faces share enough vertices to span a $(n_v+n_e-k-1)$-dimensional space. However, as we will show in the next section, this is a {\it necessary but not sufficient condition} for such a face to exist.


\section{Steinmann Relations and the Scattering Facet}\label{sec:SRSF}

One facet of particular interest is the so-called {\it scattering facet} $\mathcal{S}_{\mathcal{G}}$, on which the total external energy vanishes and is consequently conserved. The residue of $\psi_{\mathcal{G}} (x_v, y_e)$ on this boundary returns the contribution to the scattering amplitude coming from the graph $\mathcal{G}$. 
We begin by investigating how the Steinmann relations are encoded in the canonical form of $\mathcal{S}_{\mathcal{G}}$, by studying the correspondence between its facet structure and cut integrals.

The scattering facet of the cosmological polytope $\mathcal{P}_{\mathcal{G}}$ is characterized by the dual vector in equation~\eqref{eq:dual_vec} when $\mathfrak{g}$ is chosen to be the full (reduced) graph $\mathcal{G}$, namely 
\begin{equation}
\mathcal{S}_{\mathcal{G}} = \mathcal{P}_{\mathcal{G}} \cap \mathcal{W}_{\mathcal{G}}\, .
\end{equation}
This facet is given by the convex hull of the $2n_e$ vertices $\{\mathbf{x}_i+\mathbf{y}_{ij}-\mathbf{x}_j,\,-\mathbf{x}_i+\mathbf{y}_{ij}+\mathbf{x}_j\}$, where we have switched to a notation in which $y_{ij}$ represents the edge between sites $x_i$ and $x_j$.
Since it corresponds to a codimension-one face of the cosmological polytope, it has dimension $n_v+n_e-2$. 

\vspace{.14cm}
\noindent {\bf Codimension-one Faces and Individual Cuts}
\vspace{.1cm}

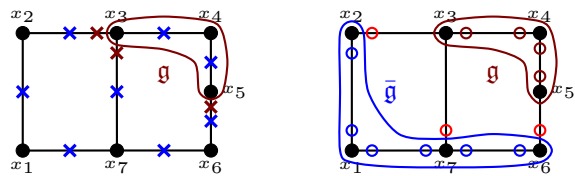
\begin{figure}[t]
\begin{tikzpicture}[ball/.style = {circle, draw, align=center, anchor=north, inner sep=0}, cross/.style={cross out, draw, minimum size=2*(#1-\pgflinewidth), inner sep=0pt, outer sep=0pt}, scale=1.25, transform shape]
  \begin{scope}
   \coordinate[label=below:{\tiny $x_1$}] (v1) at (0,0);
   \coordinate[label=above:{\tiny $x_2$}] (v2) at ($(v1)+(0,1.25)$);
   \coordinate[label=above:{\tiny $x_3$}] (v3) at ($(v2)+(1,0)$);
   \coordinate[label=above:{\tiny $x_4$}] (v4) at ($(v3)+(1,0)$);
   \coordinate[label=right:{\tiny $x_5$}] (v5) at ($(v4)-(0,.625)$);
   \coordinate[label=below:{\tiny $x_6$}] (v6) at ($(v5)-(0,.625)$);
   \coordinate[label=below:{\tiny $x_7$}] (v7) at ($(v6)-(1,0)$);
   \draw[thick] (v1) -- (v2) -- (v3) -- (v4) -- (v5) -- (v6) -- (v7) -- cycle;
   \draw[thick] (v3) -- (v7);
   \draw[fill=black] (v1) circle (2pt);
   \draw[fill=black] (v2) circle (2pt);
   \draw[fill=black] (v3) circle (2pt);
   \draw[fill=black] (v4) circle (2pt);
   \draw[fill=black] (v5) circle (2pt);
   \draw[fill=black] (v6) circle (2pt);
   \draw[fill=black] (v7) circle (2pt);
   \coordinate (v12) at ($(v1)!0.5!(v2)$);
   \coordinate (v23) at ($(v2)!0.5!(v3)$);
   \coordinate (v34) at ($(v3)!0.5!(v4)$);
   \coordinate (v45) at ($(v4)!0.5!(v5)$);
   \coordinate (v56) at ($(v5)!0.5!(v6)$);
   \coordinate (v67) at ($(v6)!0.5!(v7)$);
   \coordinate (v71) at ($(v7)!0.5!(v1)$);
   \coordinate (v37) at ($(v3)!0.5!(v7)$);
   \node[very thick, cross=4pt, rotate=0, color=blue, scale=.625] at (v12) {};
   \node[very thick, cross=4pt, rotate=0, color=blue, scale=.625] at (v23) {};
   \node[very thick, cross=4pt, rotate=0, color=blue, scale=.625] at (v34) {};
   \node[very thick, cross=4pt, rotate=0, color=blue, scale=.625] at (v45) {};
   \node[very thick, cross=4pt, rotate=0, color=blue, scale=.625] at (v56) {};
   \node[very thick, cross=4pt, rotate=0, color=blue, scale=.625] at (v67) {};
   \node[very thick, cross=4pt, rotate=0, color=blue, scale=.625] at (v71) {};
   \node[very thick, cross=4pt, rotate=0, color=blue, scale=.625] at (v37) {};
   \node[very thick, cross=4pt, rotate=0, color=red!50!black, scale=.625, left=.15cm of v3] (v3l) {};
   \node[very thick, cross=4pt, rotate=0, color=red!50!black, scale=.625, below=.15cm of v3] (v3b) {};
   \node[very thick, cross=4pt, rotate=0, color=red!50!black, scale=.625, below=.1cm of v5] (v5b){};
   \coordinate (a) at ($(v3l)!0.5!(v3)$);
   \coordinate (b) at ($(v3)+(0,.125)$);
   \coordinate (c) at ($(v34)+(0,.175)$);
   \coordinate (d) at ($(v4)+(0,.125)$);
   \coordinate (e) at ($(v4)+(.125,0)$);
   \coordinate (f) at ($(v45)+(.175,0)$);
   \coordinate (g) at ($(v5)+(.125,0)$);
   \coordinate (h) at ($(v5b)!0.5!(v5)$);
   \coordinate (i) at ($(v5)-(.125,0)$);
   \coordinate (j) at ($(v45)-(.175,0)$);
   \coordinate (k) at ($(v34)-(0,.175)$);
   \coordinate (l) at ($(v3)-(0,.125)$);
   \draw [thick, red!50!black] plot [smooth cycle] coordinates {(a) (b) (c) (d) (e) (f) (g) (h) (i) (j) (k) (l)};
   \node[below=.05cm of k, color=red!50!black] {\footnotesize $\displaystyle\mathfrak{g}$};
  \end{scope}
  \begin{scope}[shift={(3.5,0)}, transform shape]
   \coordinate[label=below:{\tiny $x_1$}] (v1) at (0,0);
   \coordinate[label=above:{\tiny $x_2$}] (v2) at ($(v1)+(0,1.25)$);
   \coordinate[label=above:{\tiny $x_3$}] (v3) at ($(v2)+(1,0)$);
   \coordinate[label=above:{\tiny $x_4$}] (v4) at ($(v3)+(1,0)$);
   \coordinate[label=right:{\tiny $x_5$}] (v5) at ($(v4)-(0,.625)$);
   \coordinate[label=below:{\tiny $x_6$}] (v6) at ($(v5)-(0,.625)$);
   \coordinate[label=below:{\tiny $x_7$}] (v7) at ($(v6)-(1,0)$);
   \draw[thick] (v1) -- (v2) -- (v3) -- (v4) -- (v5) -- (v6) -- (v7) -- cycle;
   \draw[thick] (v3) -- (v7);
   \draw[fill=black] (v1) circle (2pt);
   \draw[fill=black] (v2) circle (2pt);
   \draw[fill=black] (v3) circle (2pt);
   \draw[fill=black] (v4) circle (2pt);
   \draw[fill=black] (v5) circle (2pt);
   \draw[fill=black] (v6) circle (2pt);
   \draw[fill=black] (v7) circle (2pt);
   \coordinate (v12) at ($(v1)!0.5!(v2)$);
   \coordinate (v23) at ($(v2)!0.5!(v3)$);
   \coordinate (v34) at ($(v3)!0.5!(v4)$);
   \coordinate (v45) at ($(v4)!0.5!(v5)$);
   \coordinate (v56) at ($(v5)!0.5!(v6)$);
   \coordinate (v67) at ($(v6)!0.5!(v7)$);
   \coordinate (v71) at ($(v7)!0.5!(v1)$);
   \coordinate (v37) at ($(v3)!0.5!(v7)$);
   \node[ball,text width=.18cm,thick,color=red!50!black,right=.15cm of v3, scale=.625] {};
   \node[ball,text width=.18cm,thick,color=red!50!black,left=.15cm of v4, scale=.625] {};
   \node[ball,text width=.18cm,thick,color=red!50!black,below=.1cm of v4, scale=.625] {};
   \node[ball,text width=.18cm,thick,color=red!50!black,above=.1cm of v5, scale=.625] {};
   \node[ball,text width=.18cm,thick,color=red,right=.15cm of v2, scale=.625] {};
   \node[ball,text width=.18cm,thick,color=red,above=.15cm of v7, scale=.625] {};
   \node[ball,text width=.18cm,thick,color=red,above=.15cm of v6, scale=.625] {};
   \node[ball,text width=.18cm,thick,color=blue,below=.15cm of v2, scale=.625] {};
   \node[ball,text width=.18cm,thick,color=blue,above=.15cm of v1, scale=.625] {};
   \node[ball,text width=.18cm,thick,color=blue,right=.15cm of v1, scale=.625] {};
   \node[ball,text width=.18cm,thick,color=blue,left=.15cm of v7, scale=.625] {};
   \node[ball,text width=.18cm,thick,color=blue,right=.15cm of v7, scale=.625] {};
   \node[ball,text width=.18cm,thick,color=blue,left=.15cm of v6, scale=.625] {};
   \coordinate (a) at ($(v3)-(.125,0)$);
   \coordinate (b) at ($(v3)+(0,.125)$);
   \coordinate (c) at ($(v34)+(0,.175)$);
   \coordinate (d) at ($(v4)+(0,.125)$);
   \coordinate (e) at ($(v4)+(.125,0)$);
   \coordinate (f) at ($(v45)+(.175,0)$);
   \coordinate (g) at ($(v5)+(.125,0)$);
   \coordinate (h) at ($(v5)-(0,.125)$);
   \coordinate (i) at ($(v5)-(.125,0)$);
   \coordinate (j) at ($(v45)-(.175,0)$);
   \coordinate (k) at ($(v34)-(0,.175)$);
   \coordinate (l) at ($(v3)-(0,.125)$);
   \draw [thick, red!50!black] plot [smooth cycle] coordinates {(a) (b) (c) (d) (e) (f) (g) (h) (i) (j) (k) (l)};
   \node[below=.05cm of k, color=red!50!black] {\footnotesize $\displaystyle\mathfrak{g}$};
   \coordinate (n) at ($(v2)+(0,.125)$);
   \coordinate (o) at ($(v2)+(.125,0)$);
   \coordinate (p) at ($(v12)+(.175,0)$);
   \coordinate (q) at ($(v71)+(0,.175)$);
   \coordinate (r) at ($(v7)+(0,.125)$);
   \coordinate (t) at ($(v67)+(0,.175)$);
   \coordinate (ta) at ($(v6)+(0,.125)$);
   \coordinate (tb) at ($(v6)+(.125,0)$);
   \coordinate (tc) at ($(v6)-(0,.125)$);
   \coordinate (td) at ($(v67)-(0,.175)$);
   \coordinate (u) at ($(v71)-(0,.175)$);
   \coordinate (w) at ($(v1)-(0,.125)$);
   \coordinate (x) at ($(v1)-(.125,0)$);
   \coordinate (y) at ($(v12)-(.125,0)$);
   \coordinate (z) at ($(v2)-(.125,0)$);
   \draw [thick, blue] plot [smooth cycle] coordinates {(n) (o) (p) (q) (r) (t) (ta) (tb) (tc) (td) (u) (w) (x) (y) (z)};
   \node[right=.05cm of p, color=blue] {\footnotesize $\displaystyle\bar{\mathfrak{g}}$};
  \end{scope}
 \end{tikzpicture}
\caption{An example of a codimension-one face of the scattering facet, which is associated with an individual cut of the scattering amplitude. 
In the diagram on the left, we mark the vertices that do not appear on the scattering facet $\mathcal{S}_{\mathcal{G}}$ by \bluecross, and the additional vertices that get eliminated when we intersect it with the facet corresponding to $\mathfrak{g}$ by \purplecross.
In the diagram on the right, we depict the vertices which {\it do} contribute to $\mathcal{P}_{\mathcal{G}} \cap \mathcal{W}_{\mathfrak{g}}$ by open circles. The vertices marked by \purplecircle\ and \bluecircle\ are associated with the scattering facets $\mathcal{S}_{\mathfrak{g}}$ and $\mathcal{S}_{\bar{\mathfrak{g}}}$, respectively. The remaining vertices, marked by \redcircle, are associated with the simplex $\Sigma_{\centernot{\mathcal{E}}}$.}
\label{fig:individual_cut}
\end{figure}

Let us first review how the codimension-one faces of $\mathcal{S}_{\mathcal{G}}$ relate to the individual cuts of $\psi_{\mathcal{G}} (x_v, y_e)$~\cite{Arkani-Hamed:2018bjr}. Each of these faces corresponds to a connected subgraph $\mathfrak{g} \subset \mathcal{G}$ and is given by the convex hull of the vertices $\{\mathbf{x}_j+\mathbf{y}_{ij}-\mathbf{x}_i | x_i \notin \mathfrak{g}  \lor x_j \in \mathfrak{g}\}$. This corresponds to deleting just the vertices of $\mathcal{S}_{\mathcal{G}}$ whose markings appear closest to $\mathfrak{g}$ on the edges that depart this subgraph. On this face, $\Omega(\mathcal{Y},\mathcal{S}_{\mathcal{G}})$ factorises into a pair of lower-dimensional scattering facets and a simplex that encodes the Lorentz-invariant phase space of the cut propagators. That is,
\begin{equation}\label{eq:CutR}
 \Omega(\mathcal{Y},\mathcal{P}_{\mathcal{G}} \cap \mathcal{W}_{\mathfrak{g}})\:=\:
  \Omega(\mathcal{Y}_{\mathfrak{g}},\mathcal{S}_{\mathfrak{g}})\times\Omega(\mathcal{Y}_{\bar{\mathfrak{g}}},\mathcal{S}_{\bar{\mathfrak{g}}})\times\Omega(\mathcal{Y}_{\centernot{\mathcal{E}}},\Sigma_{\centernot{\mathcal{E}}}),
\end{equation}
where $\mathcal{S}_{\mathfrak{g}}$ and $\mathcal{S}_{\mathfrak{\bar{g}}}$ are the scattering facets associated with the reduced graph $\mathfrak{g}$ and its complement $\mathfrak{\bar{g}}$, and $\Sigma_{\centernot{\mathcal{E}}}$ is the simplex formed out of the remaining vertices of the scattering facet on the cut edges $\centernot{\mathcal{E}}$~\cite{Arkani-Hamed:2018bjr}. The direction of energy flow is encoded by the vertex structure of $\Sigma_{\centernot{\mathcal{E}}}$. An example is shown in Figure~\ref{fig:individual_cut}.

We highlight that, since we have only computed a single cut, the vertices of this face must span a space of dimension $n_v+n_e-3$. Thus, we must have
\begin{equation}\label{eq:gGcond}
2n_e-n_{\mathfrak{g}}\:\ge\:n_v+n_e-2\:=\:2n_e-L-1\, ,
\end{equation}
where $n_{\mathfrak{g}}$ is the number of edges departing from $\mathfrak{g}$, and the last equality is obtained using the fact that $n_v=n_e-L+1$, where $L$ indicates the loop order of $\mathcal{G}$. This can be rephrased as
\begin{equation}\label{eq:gGcond2}
1\:\le\:n_{\mathfrak{g}}\:\le\:L+1\,.
\end{equation}
This, it turns out, constitutes both a necessary and sufficient condition for this codimension-one face to exist, and for the corresponding cut to be nonzero.

\vspace{.14cm}
\noindent {\bf Codimension-two Faces and Sequential Cuts}
\vspace{.1cm}

We now consider the intersection of pairs of codimension-one faces on the scattering facet, which correspond to sequential cuts of the scattering amplitude. We will in particular be interested in pairs of subgraphs $\mathfrak{g}_1$ and $\mathfrak{g}_2$ that satisfy 
\begin{align}\label{eq:partially_overlapping_graphs}
\begin{split} 
    \mathfrak{g}_1 \cap \mathfrak{g}_2 \neq &\varnothing, \quad  \mathfrak{g}_1 \cap \bar{\mathfrak{g}}_2 \neq \varnothing\, , \\
    \bar{\mathfrak{g}}_1 \cap \mathfrak{g}_2 \neq &\varnothing, \quad
    \bar{\mathfrak{g}}_1 \cap \bar{\mathfrak{g}}_2 \neq \varnothing \, ,
\end{split}
\end{align}
where $\bar{\mathfrak{g}}_j$ denotes the complement of $\mathfrak{g}_j$. Such a pair of subgraphs corresponds to a pair of partially-overlapping momentum channels; sequential cuts in these channels are therefore expected to vanish by virtue of the Steinmann relations.

Sequential cuts can be nonzero only if the corresponding codimension-two facet of $\mathcal{S}_{\mathcal{G}}$ exists, and this facet can only exist if there are a sufficiently large number of vertices in this intersection to span a space of dimension $n_v + n_e - 4 = 2n_e-L-3$. This corresponds to the requirement that
\begin{equation}\label{eq:SteinRel1}
2n_e-n_{\mathfrak{g}_1}-n_{\mathfrak{g}_2}+n_{\mathfrak{g}_1\cap\mathfrak{g}_2}\:\ge\:2n_e-L-2 \, ,
\end{equation}
where $n_{\mathfrak{g}_j}$ denotes the number of edges departing from $\mathfrak{g}_j$ and $n_{\mathfrak{g}_1\cap\mathfrak{g}_2}$ is the number of edges simultaneously departing from $\mathfrak{g}_1$ and $\mathfrak{g}_2$.\footnote{To be more precise, \(n_{\mathfrak{g}}\) is the sum of the degrees of the vertices in \(\mathfrak{g}\) with respect to edges not in \(\mathfrak{g}\).  If an edge does not belong to \(\mathfrak{g}\) but \emph{both} its sites are in \(\mathfrak{g}\), then it contributes \(2\) to \(n_{\mathfrak{g}}\), even though it is a \emph{single} edge.} This inequality can be rewritten as
\begin{equation}\label{eq:SteinRel2}
 n_{\mathfrak{g}_1}+n_{\mathfrak{g}_2}-n_{\mathfrak{g}_1\cap\mathfrak{g}_2}\:\le\:L+2.
\end{equation}
However, unlike the inequality that must be satisfied by codimension-one faces of $\mathcal{S}_{\mathcal{G}}$ in equation~\eqref{eq:gGcond2}, this turns out to be a {\it necessary but not sufficient} condition for a codimension-two facet to exist. 

To see this, let us consider the general form this intersection of codimension-one facets will take. Just as the canonical function factorized into a pair of scattering facets on the intersection of $\mathcal{S}_{\mathcal{G}}$ with one of its codimension-one faces, the canonical function will now factorize into up to four scattering facets, associated with the subgraphs $\mathfrak{g}_1\cap\mathfrak{g}_2$, $\mathfrak{g}_1 \cap \bar{\mathfrak{g}}_2$, $\mathfrak{g}_2 \cap \bar{\mathfrak{g}}_1$, and $\bar{\mathfrak{g}}_1\cap\bar{\mathfrak{g}}_2$.\footnote{Note that some of these scattering facets may no longer correspond to subgraphs that are connected.} In some cases, one of these four subgraphs will be empty, and the canonical form will factorize into only three scattering facets. 
There will also in general be contributions coming from a simplex $\Sigma_{\centernot{\mathcal{E}}}$, 
corresponding to the cut edges among $\mathfrak{g}_1\cap\mathfrak{g}_2$, $\mathfrak{g}_1 \cap \bar{\mathfrak{g}}_2$, $\mathfrak{g}_2 \cap \bar{\mathfrak{g}}_1$, and
$\bar{\mathfrak{g}}_1\cap\bar{\mathfrak{g}}_2$.

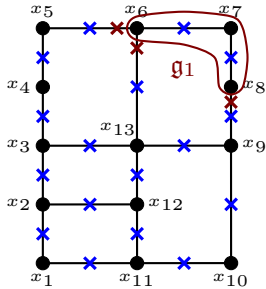
\begin{figure}
 \centering
 \vspace{-.5cm}
 \begin{tikzpicture}[ball/.style = {circle, draw, align=center, anchor=north, inner sep=0}, cross/.style={cross out, draw, minimum size=2*(#1-\pgflinewidth), inner sep=0pt, outer sep=0pt}, scale=1.25, transform shape]
  \begin{scope}
   \coordinate[label=left:{\tiny $x_3$}] (v1) at (0,0);
   \coordinate[label=above:{\tiny $x_5$}] (v2) at ($(v1)+(0,1.25)$);
   \coordinate[label=above:{\tiny $x_6$}] (v3) at ($(v2)+(1,0)$);
   \coordinate[label=above:{\tiny $x_7$}] (v4) at ($(v3)+(1,0)$);
   \coordinate[label=right:{\tiny $x_8$}] (v5) at ($(v4)-(0,.625)$);
   \coordinate[label=right:{\tiny $x_9$}] (v6) at ($(v5)-(0,.625)$);
   \coordinate[label={{[shift={(-0.2,0)}]\tiny $x_{13}$}}] (v7) at ($(v6)-(1,0)$);
   \coordinate[label=below:{\tiny $x_1$}] (v8) at ($(v1)-(0,1.25)$);
   \coordinate[label=below:{\tiny $x_{11}$}] (v9) at ($(v7)-(0,1.25)$);
   \coordinate[label=below:{\tiny $x_{10}$}] (v10) at ($(v6)-(0,1.25)$);
   \coordinate[label=left:{\tiny $x_{2}$}] (v11) at ($(v1)!0.5!(v8)$);
   \coordinate[label=right:{\tiny $x_{12}$}] (v12) at ($(v7)!0.5!(v9)$);
   \coordinate[label=left:{\tiny $x_{4}$}] (v13) at ($(v1)!0.5!(v2)$);   
   \draw[thick] (v2) -- (v4) -- (v10) -- (v8)  -- cycle;
   \draw[thick] (v1) -- (v6);   
   \draw[thick] (v3) -- (v9);
   \draw[thick] (v11) -- (v12);   
   \draw[fill=black] (v1) circle (2pt);
   \draw[fill=black] (v2) circle (2pt);
   \draw[fill=black] (v3) circle (2pt);
   \draw[fill=black] (v4) circle (2pt);
   \draw[fill=black] (v5) circle (2pt);
   \draw[fill=black] (v6) circle (2pt);
   \draw[fill=black] (v7) circle (2pt);
   \draw[fill=black] (v8) circle (2pt);
   \draw[fill=black] (v9) circle (2pt);
   \draw[fill=black] (v10) circle (2pt);
   \draw[fill=black] (v11) circle (2pt);
   \draw[fill=black] (v12) circle (2pt);
   \draw[fill=black] (v13) circle (2pt);
   \coordinate (v113) at ($(v1)!0.5!(v13)$);
   \coordinate (v132) at ($(v2)!0.5!(v13)$);
   \coordinate (v23) at ($(v2)!0.5!(v3)$);
   \coordinate (v34) at ($(v3)!0.5!(v4)$);
   \coordinate (v45) at ($(v4)!0.5!(v5)$);
   \coordinate (v56) at ($(v5)!0.5!(v6)$);
   \coordinate (v67) at ($(v6)!0.5!(v7)$);
   \coordinate (v71) at ($(v7)!0.5!(v1)$);
   \coordinate (v37) at ($(v3)!0.5!(v7)$);
   \coordinate (v111) at ($(v1)!0.5!(v11)$);
   \coordinate (v118) at ($(v11)!0.5!(v8)$);
   \coordinate (v712) at ($(v7)!0.5!(v12)$);
   \coordinate (v129) at ($(v12)!0.5!(v9)$);
   \coordinate (v610) at ($(v6)!0.5!(v10)$);
   \coordinate (v89) at ($(v8)!0.5!(v9)$);
   \coordinate (v910) at ($(v9)!0.5!(v10)$);
   \coordinate (v1112) at ($(v11)!0.5!(v12)$);
   \node[very thick, cross=4pt, rotate=0, color=blue, scale=.625] at (v113) {};
   \node[very thick, cross=4pt, rotate=0, color=blue, scale=.625] at (v132) {};
   \node[very thick, cross=4pt, rotate=0, color=blue, scale=.625] at (v23) {};
   \node[very thick, cross=4pt, rotate=0, color=blue, scale=.625] at (v34) {};
   \node[very thick, cross=4pt, rotate=0, color=blue, scale=.625] at (v45) {};
   \node[very thick, cross=4pt, rotate=0, color=blue, scale=.625] at (v56) {};
   \node[very thick, cross=4pt, rotate=0, color=blue, scale=.625] at (v67) {};
   \node[very thick, cross=4pt, rotate=0, color=blue, scale=.625] at (v71) {};
   \node[very thick, cross=4pt, rotate=0, color=blue, scale=.625] at (v37) {};
   \node[very thick, cross=4pt, rotate=0, color=blue, scale=.625] at (v111) {};
   \node[very thick, cross=4pt, rotate=0, color=blue, scale=.625] at (v118) {};
   \node[very thick, cross=4pt, rotate=0, color=blue, scale=.625] at (v712) {};
   \node[very thick, cross=4pt, rotate=0, color=blue, scale=.625] at (v129) {};
   \node[very thick, cross=4pt, rotate=0, color=blue, scale=.625] at (v610) {};
   \node[very thick, cross=4pt, rotate=0, color=blue, scale=.625] at (v89) {};
   \node[very thick, cross=4pt, rotate=0, color=blue, scale=.625] at (v910) {};
   \node[very thick, cross=4pt, rotate=0, color=blue, scale=.625] at (v1112) {};
   \node[very thick, cross=4pt, rotate=0, color=red!50!black, scale=.625, left=.15cm of v3] (v3l) {};
   \node[very thick, cross=4pt, rotate=0, color=red!50!black, scale=.625, below=.15cm of v3] (v3b) {};
   \node[very thick, cross=4pt, rotate=0, color=red!50!black, scale=.625, below=.1cm of v5] (v5b){};
   \coordinate (a) at ($(v3l)!0.5!(v3)$);
   \coordinate (b) at ($(v3)+(0,.125)$);
   \coordinate (c) at ($(v34)+(0,.175)$);
   \coordinate (d) at ($(v4)+(0,.125)$);
   \coordinate (e) at ($(v4)+(.125,0)$);
   \coordinate (f) at ($(v45)+(.175,0)$);
   \coordinate (g) at ($(v5)+(.125,0)$);
   \coordinate (h) at ($(v5b)!0.5!(v5)$);
   \coordinate (i) at ($(v5)-(.125,0)$);
   \coordinate (j) at ($(v45)-(.175,0)$);
   \coordinate (k) at ($(v34)-(0,.175)$);
   \coordinate (l) at ($(v3)-(0,.125)$);
   \draw [thick, red!50!black] plot [smooth cycle] coordinates {(a) (b) (c) (d) (e) (f) (g) (h) (i) (j) (k) (l)};
   \node[below=.05cm of k, color=red!50!black] {\footnotesize $\displaystyle\mathfrak{g}_1$};
  \end{scope}
  \begin{scope}[shift={(3.5,0)}, transform shape]
   \coordinate[label=left:{\tiny $x_3$}] (v1) at (0,0);
   \coordinate[label=above:{\tiny $x_5$}] (v2) at ($(v1)+(0,1.25)$);
   \coordinate[label=above:{\tiny $x_6$}] (v3) at ($(v2)+(1,0)$);
   \coordinate[label=above:{\tiny $x_7$}] (v4) at ($(v3)+(1,0)$);
   \coordinate[label=right:{\tiny $x_8$}] (v5) at ($(v4)-(0,.625)$);
   \coordinate[label=right:{\tiny $x_9$}] (v6) at ($(v5)-(0,.625)$);
   \coordinate[label={{[shift={(-0.2,0)}]\tiny $x_{13}$}}] (v7) at ($(v6)-(1,0)$);
   \coordinate[label=below:{\tiny $x_1$}] (v8) at ($(v1)-(0,1.25)$);
   \coordinate[label=below:{\tiny $x_{11}$}] (v9) at ($(v7)-(0,1.25)$);
   \coordinate[label=below:{\tiny $x_{10}$}] (v10) at ($(v6)-(0,1.25)$);
   \coordinate[label=left:{\tiny $x_{2}$}] (v11) at ($(v1)!0.5!(v8)$);
   \coordinate[label=right:{\tiny $x_{12}$}] (v12) at ($(v7)!0.5!(v9)$);
   \coordinate[label=left:{\tiny $x_{4}$}] (v13) at ($(v1)!0.5!(v2)$);   
   \draw[thick] (v2) -- (v4) -- (v10) -- (v8)  -- cycle;
   \draw[thick] (v1) -- (v6);   
   \draw[thick] (v3) -- (v9);
   \draw[thick] (v11) -- (v12);   
   \draw[fill=black] (v1) circle (2pt);
   \draw[fill=black] (v2) circle (2pt);
   \draw[fill=black] (v3) circle (2pt);
   \draw[fill=black] (v4) circle (2pt);
   \draw[fill=black] (v5) circle (2pt);
   \draw[fill=black] (v6) circle (2pt);
   \draw[fill=black] (v7) circle (2pt);
   \draw[fill=black] (v8) circle (2pt);
   \draw[fill=black] (v9) circle (2pt);
   \draw[fill=black] (v10) circle (2pt);
   \draw[fill=black] (v11) circle (2pt);
   \draw[fill=black] (v12) circle (2pt);
   \draw[fill=black] (v13) circle (2pt);
   \coordinate (v113) at ($(v1)!0.5!(v13)$);
   \coordinate (v132) at ($(v2)!0.5!(v13)$);
   \coordinate (v23) at ($(v2)!0.5!(v3)$);
   \coordinate (v34) at ($(v3)!0.5!(v4)$);
   \coordinate (v45) at ($(v4)!0.5!(v5)$);
   \coordinate (v56) at ($(v5)!0.5!(v6)$);
   \coordinate (v67) at ($(v6)!0.5!(v7)$);
   \coordinate (v71) at ($(v7)!0.5!(v1)$);
   \coordinate (v37) at ($(v3)!0.5!(v7)$);
   \coordinate (v111) at ($(v1)!0.5!(v11)$);
   \coordinate (v118) at ($(v11)!0.5!(v8)$);
   \coordinate (v712) at ($(v7)!0.5!(v12)$);
   \coordinate (v129) at ($(v12)!0.5!(v9)$);
   \coordinate (v610) at ($(v6)!0.5!(v10)$);
   \coordinate (v89) at ($(v8)!0.5!(v9)$);
   \coordinate (v910) at ($(v9)!0.5!(v10)$);
   \coordinate (v1112) at ($(v11)!0.5!(v12)$);
   \coordinate (v1r) at ($(v1)+(.15,0)$);
   \coordinate (v1d) at ($(v1)-(0,.15)$);
   \coordinate (v3d) at ($(v3)-(0,.15)$);
   \coordinate (v4d) at ($(v4)-(0,.15)$);
   \node[very thick, cross=4pt, rotate=0, color=blue, scale=.625] at (v113) {};   
   \node[very thick, cross=4pt, rotate=0, color=blue, scale=.625] at (v132) {};   
   \node[very thick, cross=4pt, rotate=0, color=blue, scale=.625] at (v23) {};   
   \node[very thick, cross=4pt, rotate=0, color=blue, scale=.625] at (v34) {};   
   \node[very thick, cross=4pt, rotate=0, color=blue, scale=.625] at (v45) {}; 
   \node[very thick, cross=4pt, rotate=0, color=blue, scale=.625] at (v56) {}; 
   \node[very thick, cross=4pt, rotate=0, color=blue, scale=.625] at (v67) {}; 
   \node[very thick, cross=4pt, rotate=0, color=blue, scale=.625] at (v71) {};
   \node[very thick, cross=4pt, rotate=0, color=blue, scale=.625] at (v37) {};
   \node[very thick, cross=4pt, rotate=0, color=blue, scale=.625] at (v111) {};
   \node[very thick, cross=4pt, rotate=0, color=blue, scale=.625] at (v118) {};
   \node[very thick, cross=4pt, rotate=0, color=blue, scale=.625] at (v712) {};
   \node[very thick, cross=4pt, rotate=0, color=blue, scale=.625] at (v129) {};
   \node[very thick, cross=4pt, rotate=0, color=blue, scale=.625] at (v610) {};
   \node[very thick, cross=4pt, rotate=0, color=blue, scale=.625] at (v89) {};
   \node[very thick, cross=4pt, rotate=0, color=blue, scale=.625] at (v910) {};
   \node[very thick, cross=4pt, rotate=0, color=blue, scale=.625] at (v1112) {};
   \node[very thick, cross=4pt, rotate=0, color=green!50!black, scale=.625] at (v1r) {};  
   \node[very thick, cross=4pt, rotate=0, color=green!50!black, scale=.625] at (v1d) {};    
   \node[very thick, cross=4pt, rotate=0, color=green!50!black, scale=.625] at (v3d) {};   
   \node[very thick, cross=4pt, rotate=0, color=green!50!black, scale=.625] at (v4d) {};
   \coordinate (a) at ($(v4)+(.125,0)$);
   \coordinate (b) at ($(v4)+(0,.125)$);
   \coordinate (c) at ($(v3)+(0,.125)$);
   \coordinate (d) at ($(v2)+(0,.125)$);
   \coordinate (e) at ($(v2)-(.125,0)$);
   \coordinate (f) at ($(v13)-(.125,0)$);
   \coordinate (g) at ($(v1)-(.125,0)$);
   \coordinate (gh) at ($(v1)-(0,.125)$);
   \coordinate (h) at ($(v1r)!0.5!(v1)$);
   \coordinate (j) at ($(v13)+(.125,0)$);
   \coordinate (k) at ($(v23)-(0,.125)$);
   \coordinate (l) at ($(v3)-(0,.125)$);
   \coordinate (m) at ($(v4)-(0,.125)$);
   \draw [thick, green!50!black] plot [smooth cycle] coordinates {(a) (b) (c) (d) (e) (f) (g) (gh) (h) (j) (k) (l) (m)};
   \node[color=green!50!black] at ($(v13)+(.5,0)$) {\footnotesize $\displaystyle\mathfrak{g}_2$};
  \end{scope}
 \end{tikzpicture}
\caption{An example of a pair of subgraphs that correspond to partially overlapping momentum channels, and their realization as faces on the scattering facet.}
\label{fig:ovch}
\end{figure}

\begin{figure}
 \centering
 \vspace{-.25cm}
 \begin{tikzpicture}[ball/.style = {circle, draw, align=center, anchor=north, inner sep=0}, cross/.style={cross out, draw, minimum size=2*(#1-\pgflinewidth), inner sep=0pt, outer sep=0pt}, scale=1.25, transform shape]
  \begin{scope}
   \coordinate[label=left:{\tiny $x_3$}] (v1) at (0,0);
   \coordinate[label=above:{\tiny $x_5$}] (v2) at ($(v1)+(0,1.25)$);
   \coordinate[label=above:{\tiny $x_6$}] (v3) at ($(v2)+(1,0)$);
   \coordinate[label=above:{\tiny $x_7$}] (v4) at ($(v3)+(1,0)$);
   \coordinate[label=right:{\tiny $x_8$}] (v5) at ($(v4)-(0,.625)$);
   \coordinate[label=right:{\tiny $x_9$}] (v6) at ($(v5)-(0,.625)$);
   \coordinate[label={{[shift={(-0.2,0)}]\tiny $x_{13}$}}] (v7) at ($(v6)-(1,0)$);
   \coordinate[label=below:{\tiny $x_1$}] (v8) at ($(v1)-(0,1.25)$);
   \coordinate[label=below:{\tiny $x_{11}$}] (v9) at ($(v7)-(0,1.25)$);
   \coordinate[label=below:{\tiny $x_{10}$}] (v10) at ($(v6)-(0,1.25)$);
   \coordinate[label=left:{\tiny $x_{2}$}] (v11) at ($(v1)!0.5!(v8)$);
   \coordinate[label=right:{\tiny $x_{12}$}] (v12) at ($(v7)!0.5!(v9)$);
   \coordinate[label=left:{\tiny $x_{4}$}] (v13) at ($(v1)!0.5!(v2)$);   
   \draw[thick] (v2) -- (v4) -- (v10) -- (v8)  -- cycle;
   \draw[thick] (v1) -- (v6);   
   \draw[thick] (v3) -- (v9);
   \draw[thick] (v11) -- (v12);   
   \draw[fill=black] (v1) circle (2pt);
   \draw[fill=black] (v2) circle (2pt);
   \draw[fill=black] (v3) circle (2pt);
   \draw[fill=black] (v4) circle (2pt);
   \draw[fill=black] (v5) circle (2pt);
   \draw[fill=black] (v6) circle (2pt);
   \draw[fill=black] (v7) circle (2pt);
   \draw[fill=black] (v8) circle (2pt);
   \draw[fill=black] (v9) circle (2pt);
   \draw[fill=black] (v10) circle (2pt);
   \draw[fill=black] (v11) circle (2pt);
   \draw[fill=black] (v12) circle (2pt);
   \draw[fill=black] (v13) circle (2pt);
   \coordinate (v113) at ($(v1)!0.5!(v13)$);
   \coordinate (v132) at ($(v2)!0.5!(v13)$);
   \coordinate (v23) at ($(v2)!0.5!(v3)$);
   \coordinate (v34) at ($(v3)!0.5!(v4)$);
   \coordinate (v45) at ($(v4)!0.5!(v5)$);
   \coordinate (v56) at ($(v5)!0.5!(v6)$);
   \coordinate (v67) at ($(v6)!0.5!(v7)$);
   \coordinate (v71) at ($(v7)!0.5!(v1)$);
   \coordinate (v37) at ($(v3)!0.5!(v7)$);
   \coordinate (v111) at ($(v1)!0.5!(v11)$);
   \coordinate (v118) at ($(v11)!0.5!(v8)$);
   \coordinate (v712) at ($(v7)!0.5!(v12)$);
   \coordinate (v129) at ($(v12)!0.5!(v9)$);
   \coordinate (v610) at ($(v6)!0.5!(v10)$);
   \coordinate (v89) at ($(v8)!0.5!(v9)$);
   \coordinate (v910) at ($(v9)!0.5!(v10)$);
   \coordinate (v1112) at ($(v11)!0.5!(v12)$);
   \coordinate (v1r) at ($(v1)+(.15,0)$);
   \coordinate (v1d) at ($(v1)-(0,.15)$);
   \coordinate (v3d) at ($(v3)-(0,.15)$);
   \coordinate (v3l) at ($(v3)-(.15,0)$);
   \coordinate (v4d) at ($(v4)-(0,.15)$);
   \coordinate (v5b) at ($(v5)-(0,.15)$);
   \coordinate (a) at ($(v3l)!0.5!(v3)$);
   \coordinate (b) at ($(v3)+(0,.125)$);
   \coordinate (c) at ($(v34)+(0,.175)$);
   \coordinate (d) at ($(v4)+(0,.125)$);
   \coordinate (e) at ($(v4)+(.125,0)$);
   \coordinate (f) at ($(v45)+(.175,0)$);
   \coordinate (g) at ($(v5)+(.125,0)$);
   \coordinate (h) at ($(v5b)!0.5!(v5)$);
   \coordinate (i) at ($(v5)-(.125,0)$);
   \coordinate (j) at ($(v45)-(.175,0)$);
   \coordinate (k) at ($(v34)-(0,.175)$);
   \coordinate (l) at ($(v3)-(0,.125)$);
   \draw [thick, red!50!black] plot [smooth cycle] coordinates {(a) (b) (c) (d) (e) (f) (g) (h) (i) (j) (k) (l)};
   \node[color=red!50!black] at ($(v45)-(.5,0)$) {\footnotesize $\displaystyle\mathfrak{g}_1$};   
   \coordinate (a2) at ($(v4)+(.125,0)$);
   \coordinate (b2) at ($(v4)+(0,.125)$);
   \coordinate (c2) at ($(v3)+(0,.125)$);
   \coordinate (d2) at ($(v2)+(0,.125)$);
   \coordinate (e2) at ($(v2)-(.125,0)$);
   \coordinate (f2) at ($(v13)-(.125,0)$);
   \coordinate (g2) at ($(v1)-(.125,0)$);
   \coordinate (gh2) at ($(v1)-(0,.125)$);
   \coordinate (h2) at ($(v1r)!0.5!(v1)$);
   \coordinate (j2) at ($(v13)+(.125,0)$);
   \coordinate (k2) at ($(v23)-(0,.125)$);
   \coordinate (l2) at ($(v3)-(0,.125)$);
   \coordinate (m2) at ($(v4)-(0,.125)$);
   \draw [thick, green!50!black] plot [smooth cycle] coordinates {(a2) (b2) (c2) (d2) (e2) (f2) (g2) (gh2) (h2) (j2) (k2) (l2) (m2)};
   \node[color=green!50!black] at ($(v13)+(.5,0)$) {\footnotesize $\displaystyle\mathfrak{g}_2$};   
   \node[ball,text width=.18cm,thick,color=red!50!green,right=.15cm of v3, scale=.625] {};
   \node[ball,text width=.18cm,thick,color=red!50!green,left=.15cm of v4, scale=.625] {};
   \node[ball,text width=.18cm,thick,color=green!50!black,right=.15cm of v2, scale=.625] {};
   \node[ball,text width=.18cm,thick,color=green!50!black,below=.075cm of v2, scale=.625] {};
   \node[ball,text width=.18cm,thick,color=green!50!black,above=.075cm of v13, scale=.625] {};
   \node[ball,text width=.18cm,thick,color=green!50!black,below=.075cm of v13, scale=.625] {};
   \node[ball,text width=.18cm,thick,color=green!50!black,above=.075cm of v1, scale=.625] {};
   \node[ball,text width=.18cm,thick,color=red!50!black,above=.075cm of v5, scale=.625] {};
   \node[ball,text width=.18cm,thick,color=red,above=.1cm of v11, scale=.625] {};
   \node[ball,text width=.18cm,thick,color=red,above=.15cm of v7, scale=.625] {};
   \node[ball,text width=.18cm,thick,color=red,left=.15cm of v7, scale=.625] {};
   \node[ball,text width=.18cm,thick,color=red,above=.15cm of v6, scale=.625] {};
   \node[ball,text width=.18cm,thick,color=blue,below=.075cm of v7, scale=.625] {};
   \node[ball,text width=.18cm,thick,color=blue,right=.15cm of v7, scale=.625] {};
   \node[ball,text width=.18cm,thick,color=blue,left=.15cm of v6, scale=.625] {};
   \node[ball,text width=.18cm,thick,color=blue,below=.15cm of v6, scale=.625] {};
   \node[ball,text width=.18cm,thick,color=blue,above=.075cm of v12, scale=.625] {};
   \node[ball,text width=.18cm,thick,color=blue,left=.15cm of v12, scale=.625] {};
   \node[ball,text width=.18cm,thick,color=blue,below=.075cm of v12, scale=.625] {};
   \node[ball,text width=.18cm,thick,color=blue,right=.15cm of v11, scale=.625] {};
   \node[ball,text width=.18cm,thick,color=blue,below=.075cm of v11, scale=.625] {};
   \node[ball,text width=.18cm,thick,color=blue,above=.075cm of v8, scale=.625] {};
   \node[ball,text width=.18cm,thick,color=blue,right=.15cm of v8, scale=.625] {};
   \node[ball,text width=.18cm,thick,color=blue,left=.15cm of v9, scale=.625] {};
   \node[ball,text width=.18cm,thick,color=blue,above=.075cm of v9, scale=.625] {};
   \node[ball,text width=.18cm,thick,color=blue,right=.15cm of v9, scale=.625] {};
   \node[ball,text width=.18cm,thick,color=blue,left=.15cm of v10, scale=.625] {};
   \node[ball,text width=.18cm,thick,color=blue,above=.15cm of v10, scale=.625] {};
   \coordinate (b3) at ($(v6)+(.15,0)$);
   \coordinate (c3) at ($(v6)+(0,.1)$);
   \coordinate (d3) at ($(v67)+(0,.1)$);
   \coordinate (e3) at ($(v7)+(0,.1)$);
   \coordinate (f3) at ($(v7)-(.1,0)$);
   \coordinate (g3) at ($(v712)-(.1,0)$);
   \coordinate (h3) at ($(v12)+(-.15,.15)$);
   \coordinate (i3) at ($(v1112)+(0,.15)$);
   \coordinate (j3) at ($(v11)+(0,.1)$);
   \coordinate (k3) at ($(v11)-(.15,0)$);
   \coordinate (l3) at ($(v118)-(.15,0)$);
   \coordinate (m3) at ($(v8)-(.15,0)$);
   \coordinate (n3) at ($(v8)-(0,.15)$);
   \coordinate (o3) at ($(v89)-(0,.15)$);
   \coordinate (p3) at ($(v9)-(0,.15)$);
   \coordinate (q3) at ($(v910)-(0,.15)$);
   \coordinate (r3) at ($(v10)-(0,.15)$);
   \coordinate (s3) at ($(v10)+(.15,0)$);
   \coordinate (t3) at ($(v610)+(.15,0)$);
   \draw [thick, blue] plot [smooth cycle] coordinates {(b3) (c3) (d3) (e3) (f3) (g3) (h3) (i3) (j3) (k3) (l3) (m3) (n3) (o3) (p3) (q3) (r3) (s3) (t3)};
   \node[color=blue, scale=.75] at ($(v712)-(.5,0)$) {\footnotesize $\displaystyle\bar{\mathfrak{g}}_{1}\cap\bar{\mathfrak{g}}_2$};   
  \end{scope}
  \begin{scope}[shift={(3.5, 0)}, transform shape]
   \coordinate[label=left:{\tiny $x_3$}] (v1) at (0,0);
   \coordinate[label=above:{\tiny $x_5$}] (v2) at ($(v1)+(0,1.25)$);
   \coordinate[label=above:{\tiny $x_6$}] (v3) at ($(v2)+(1,0)$);
   \coordinate[label=above:{\tiny $x_7$}] (v4) at ($(v3)+(1,0)$);
   \coordinate[label=right:{\tiny $x_8$}] (v5) at ($(v4)-(0,.625)$);
   \coordinate[label=right:{\tiny $x_9$}] (v6) at ($(v5)-(0,.625)$);
   \coordinate[label={{[shift={(-0.2,0)}]\tiny $x_{13}$}}] (v7) at ($(v6)-(1,0)$);
   \coordinate[label=below:{\tiny $x_1$}] (v8) at ($(v1)-(0,1.25)$);
   \coordinate[label=below:{\tiny $x_{11}$}] (v9) at ($(v7)-(0,1.25)$);
   \coordinate[label=below:{\tiny $x_{10}$}] (v10) at ($(v6)-(0,1.25)$);
   \coordinate[label=left:{\tiny $x_{2}$}] (v11) at ($(v1)!0.5!(v8)$);
   \coordinate[label=right:{\tiny $x_{12}$}] (v12) at ($(v7)!0.5!(v9)$);
   \coordinate[label=left:{\tiny $x_{4}$}] (v13) at ($(v1)!0.5!(v2)$);   
   \draw[thick] (v2) -- (v4) -- (v10) -- (v8)  -- cycle;
   \draw[thick] (v1) -- (v6);   
   \draw[thick] (v3) -- (v9);
   \draw[thick] (v11) -- (v12);   
   \draw[fill=black] (v1) circle (2pt);
   \draw[fill=black] (v2) circle (2pt);
   \draw[fill=black] (v3) circle (2pt);
   \draw[fill=black] (v4) circle (2pt);
   \draw[fill=black] (v5) circle (2pt);
   \draw[fill=black] (v6) circle (2pt);
   \draw[fill=black] (v7) circle (2pt);
   \draw[fill=black] (v8) circle (2pt);
   \draw[fill=black] (v9) circle (2pt);
   \draw[fill=black] (v10) circle (2pt);
   \draw[fill=black] (v11) circle (2pt);
   \draw[fill=black] (v12) circle (2pt);
   \draw[fill=black] (v13) circle (2pt);
   \coordinate (v113) at ($(v1)!0.5!(v13)$);
   \coordinate (v132) at ($(v2)!0.5!(v13)$);
   \coordinate (v23) at ($(v2)!0.5!(v3)$);
   \coordinate (v34) at ($(v3)!0.5!(v4)$);
   \coordinate (v45) at ($(v4)!0.5!(v5)$);
   \coordinate (v56) at ($(v5)!0.5!(v6)$);
   \coordinate (v67) at ($(v6)!0.5!(v7)$);
   \coordinate (v71) at ($(v7)!0.5!(v1)$);
   \coordinate (v37) at ($(v3)!0.5!(v7)$);
   \coordinate (v111) at ($(v1)!0.5!(v11)$);
   \coordinate (v118) at ($(v11)!0.5!(v8)$);
   \coordinate (v712) at ($(v7)!0.5!(v12)$);
   \coordinate (v129) at ($(v12)!0.5!(v9)$);
   \coordinate (v610) at ($(v6)!0.5!(v10)$);
   \coordinate (v89) at ($(v8)!0.5!(v9)$);
   \coordinate (v910) at ($(v9)!0.5!(v10)$);
   \coordinate (v1112) at ($(v11)!0.5!(v12)$);
   \coordinate (v1r) at ($(v1)+(.15,0)$);
   \coordinate (v1d) at ($(v1)-(0,.15)$);
   \coordinate (v3d) at ($(v3)-(0,.15)$);
   \coordinate (v3l) at ($(v3)-(.15,0)$);
   \coordinate (v4d) at ($(v4)-(0,.15)$);
   \coordinate (v5b) at ($(v5)-(0,.15)$);
   \coordinate (a) at ($(v3)-(.125,0)$);
   \coordinate (b) at ($(v3)+(0,.125)$);
   \coordinate (c) at ($(v34)+(0,.175)$);
   \coordinate (d) at ($(v4)+(0,.125)$);
   \coordinate (e) at ($(v4)+(.125,0)$);
   \coordinate (f) at ($(v4)-(0,.125)$);
   \coordinate (k) at ($(v34)-(0,.175)$);
   \coordinate (l) at ($(v3)-(0,.125)$);
   \draw [thick, red!50!black] (v5) circle (3pt);
   \node[color=red!50!black, scale=.75] at ($(v5)-(.475,0)$) {\footnotesize $\displaystyle\mathfrak{g}_1\cap\bar{\mathfrak{g}}_2$};   
   \draw [thick, red!50!green] plot [smooth cycle] coordinates {(a) (b) (c) (d) (e) (f) (k) (l)};
   \node[color=red!50!green, scale=.75] at ($(v34)+(0,.375)$) {\footnotesize $\displaystyle\mathfrak{g}_1\cap\mathfrak{g}_2$};   
   \coordinate (a2) at ($(v2)+(.125,0)$);
   \coordinate (d2) at ($(v2)+(0,.125)$);
   \coordinate (e2) at ($(v2)-(.125,0)$);
   \coordinate (f2) at ($(v13)-(.125,0)$);
   \coordinate (g2) at ($(v1)-(.125,0)$);
   \coordinate (gh2) at ($(v1)-(0,.125)$);
   \coordinate (h2) at ($(v1)+(.125,0)$);
   \coordinate (j2) at ($(v13)+(.125,0)$);
   \draw [thick, green!50!black] plot [smooth cycle] coordinates {(a2) (d2) (e2) (f2) (g2) (gh2) (h2) (j2)};
   \node[color=green!50!black, scale=.75] at ($(v13)+(.475,0)$) {\footnotesize $\displaystyle\mathfrak{g}_2\cap\bar{\mathfrak{g}}_1$};   
   \node[ball,text width=.18cm,thick,color=red!50!green,right=.15cm of v3, scale=.625] {};
   \node[ball,text width=.18cm,thick,color=red!50!green,left=.15cm of v4, scale=.625] {};
   \node[ball,text width=.18cm,thick,color=red,right=.15cm of v2, scale=.625] {};
   \node[ball,text width=.18cm,thick,color=green!50!black,below=.075cm of v2, scale=.625] {};
   \node[ball,text width=.18cm,thick,color=green!50!black,above=.075cm of v13, scale=.625] {};
   \node[ball,text width=.18cm,thick,color=green!50!black,below=.075cm of v13, scale=.625] {};
   \node[ball,text width=.18cm,thick,color=green!50!black,above=.075cm of v1, scale=.625] {};
   \node[ball,text width=.18cm,thick,color=red,above=.1cm of v5, scale=.625] {};
   \node[ball,text width=.18cm,thick,color=red,above=.1cm of v11, scale=.625] {};
   \node[ball,text width=.18cm,thick,color=red,above=.15cm of v7, scale=.625] {};
   \node[ball,text width=.18cm,thick,color=red,left=.15cm of v7, scale=.625] {};
   \node[ball,text width=.18cm,thick,color=red,above=.15cm of v6, scale=.625] {};
   \node[ball,text width=.18cm,thick,color=blue,below=.075cm of v7, scale=.625] {};
   \node[ball,text width=.18cm,thick,color=blue,right=.15cm of v7, scale=.625] {};
   \node[ball,text width=.18cm,thick,color=blue,left=.15cm of v6, scale=.625] {};
   \node[ball,text width=.18cm,thick,color=blue,below=.15cm of v6, scale=.625] {};
   \node[ball,text width=.18cm,thick,color=blue,above=.075cm of v12, scale=.625] {};
   \node[ball,text width=.18cm,thick,color=blue,left=.15cm of v12, scale=.625] {};
   \node[ball,text width=.18cm,thick,color=blue,below=.075cm of v12, scale=.625] {};
   \node[ball,text width=.18cm,thick,color=blue,right=.15cm of v11, scale=.625] {};
   \node[ball,text width=.18cm,thick,color=blue,below=.075cm of v11, scale=.625] {};
   \node[ball,text width=.18cm,thick,color=blue,above=.075cm of v8, scale=.625] {};
   \node[ball,text width=.18cm,thick,color=blue,right=.15cm of v8, scale=.625] {};
   \node[ball,text width=.18cm,thick,color=blue,left=.15cm of v9, scale=.625] {};
   \node[ball,text width=.18cm,thick,color=blue,above=.075cm of v9, scale=.625] {};
   \node[ball,text width=.18cm,thick,color=blue,right=.15cm of v9, scale=.625] {};
   \node[ball,text width=.18cm,thick,color=blue,left=.15cm of v10, scale=.625] {};
   \node[ball,text width=.18cm,thick,color=blue,above=.15cm of v10, scale=.625] {};
   \coordinate (b3) at ($(v6)+(.15,0)$);
   \coordinate (c3) at ($(v6)+(0,.1)$);
   \coordinate (d3) at ($(v67)+(0,.1)$);
   \coordinate (e3) at ($(v7)+(0,.1)$);
   \coordinate (f3) at ($(v7)-(.1,0)$);
   \coordinate (g3) at ($(v712)-(.1,0)$);
   \coordinate (h3) at ($(v12)+(-.15,.15)$);
   \coordinate (i3) at ($(v1112)+(0,.15)$);
   \coordinate (j3) at ($(v11)+(0,.1)$);
   \coordinate (k3) at ($(v11)-(.15,0)$);
   \coordinate (l3) at ($(v118)-(.15,0)$);
   \coordinate (m3) at ($(v8)-(.15,0)$);
   \coordinate (n3) at ($(v8)-(0,.15)$);
   \coordinate (o3) at ($(v89)-(0,.15)$);
   \coordinate (p3) at ($(v9)-(0,.15)$);
   \coordinate (q3) at ($(v910)-(0,.15)$);
   \coordinate (r3) at ($(v10)-(0,.15)$);
   \coordinate (s3) at ($(v10)+(.15,0)$);
   \coordinate (t3) at ($(v610)+(.15,0)$);
   \draw [thick, blue] plot [smooth cycle] coordinates {(b3) (c3) (d3) (e3) (f3) (g3) (h3) (i3) (j3) (k3) (l3) (m3) (n3) (o3) (p3) (q3) (r3) (s3) (t3)};
   \node[color=blue, scale=.75] at ($(v712)-(.5,0)$) {\footnotesize $\displaystyle\bar{\mathfrak{g}}_{1}\cap\bar{\mathfrak{g}}_{2}$};   
  \end{scope}
 \end{tikzpicture}
\caption{The intersection of the pair of facets depicted in Figure~\ref{fig:ovch}. This intersection factorizes into four lower-dimensional scattering facets $\mathcal{S}_{\mathfrak{g}_1 \cap \mathfrak{g}_2}$, $\mathcal{S}_{\mathfrak{g}_1 \cap \bar{\mathfrak{g}}_2}$, $\mathcal{S}_{\bar{\mathfrak{g}}_1 \cap \mathfrak{g}_2}$ and $\mathcal{S}_{\bar{\mathfrak{g}}_1\cap\bar{\mathfrak{g}}_2}$, whose vertices are respectively depicted by the markings
\circle{red!50!black},
\circle{red!50!green},
\circle{green!50!black},
and
\circle{green!50!black}.
The remaining vertices, denoted by
\redcircle{}
, identify the  simplex $\Sigma_{\centernot{\mathcal{E}}}$.
This represents an example of an intersection that satisfies condition \eqref{eq:SteinRel2}, yet that does not form a codimension-two facet of $\mathcal{S}_{\mathcal{G}}$.}
 \label{fig:overlapping_channels}
\end{figure}
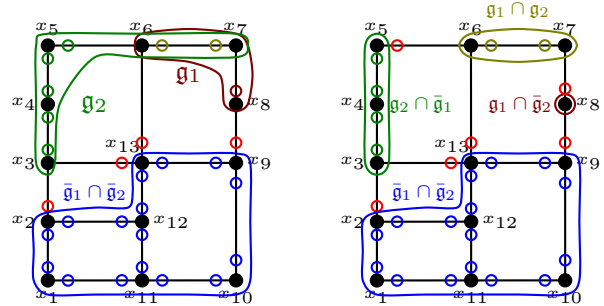

Importantly, each of these lower-dimensional polytopes lives in a space whose dimension is determined by the number of sites and edges of the associated subgraph. In particular, a scattering facet associated with a graph $\mathfrak{g}$ that has $n^{\mathfrak{g}}_{v}$ sites and $n^{\mathfrak{g}}_{e}$ edges will have dimension $n^{\mathfrak{g}}_{v}+n^{\mathfrak{g}}_{e}-2$, while $\Sigma_{\centernot{\mathcal{E}}}$
will have dimension $n_{\not{\mathcal{E}}} -1$,
where $n_{\centernot{\mathcal{E}}}$
denotes the number of cut edges in $\mathcal{G}$ associated with the simplex. Given this,
we can simply check whether these polytopes jointly give rise to a space of dimension $n_v+n_e-4$, as required.
We find
\begin{equation}\label{eq:nsc}\begin{split}
  \text{dim}(\mathcal{S}_{\mathcal{G}} \cap \mathcal{W}_{\mathfrak{g}_1} \cap \mathcal{W}_{\mathfrak{g}_2}) = \sum_{\mathcal{S}_\mathfrak{g}} \left(n_v^{\mathfrak{g}}+n_e^{\mathfrak{g}} - 1 \right) + n_{\centernot{\mathcal{E}}}
  - 1 \, ,
  \end{split}
\end{equation}
where the sum is over the scattering facets formed by the subgraphs $\mathfrak{g}_1\cap\mathfrak{g}_2$, $\mathfrak{g}_1 \cap \bar{\mathfrak{g}}_2$, $\mathfrak{g}_2 \cap \bar{\mathfrak{g}}_1$, and $\bar{\mathfrak{g}}_1\cap\bar{\mathfrak{g}}_2$. Given that all sites in $\mathcal{G}$ contribute to a single scattering facet, and that all of its edges are associated either with a scattering facet or the simplex $\Sigma_{\centernot{\mathcal{E}}}$,
this formula reduces to
\begin{equation}\label{eq:dim_intersection_2}
  \text{dim}(\mathcal{S}_{\mathcal{G}} \cap \mathcal{W}_{\mathfrak{g}_1} \cap \mathcal{W}_{\mathfrak{g}_2}) = n_v + n_e - 1 - \sum_{\mathcal{S}_\mathfrak{g}} 1 \, .
\end{equation}
Clearly this will reproduce the correct dimension only if the sum is over three scattering facets, and not four. 

We now recall that pairs of graphs $\mathfrak{g}_1$ and $\mathfrak{g}_2$ that that correspond to partially-overlapping momentum channels satisfy equation~\eqref{eq:partially_overlapping_graphs}. It is easy to see that the intersection $\mathcal{S}_{\mathcal{G}}\cap \mathcal{W}_{\mathfrak{g}_1} \cap \mathcal{W}_{\mathfrak{g}_2}$ for such subgraphs will always factorize into four nontrivial scattering facets; an example is shown in Figure~\ref{fig:overlapping_channels}. As such, it follows from equation~\eqref{eq:dim_intersection_2} that the intersection of the codimension-one faces corresponding to these subgraphs will always constitute a  codimension-three facet of $\mathcal{S}_\mathcal{G}$. This implies that these cuts must vanish, and thereby encodes the Steinmann relations in a combinatorial way. Conversely, it is easy to see that pairs of graphs $\mathfrak{g}_1$ and $\mathfrak{g}_2$ that dot no satisfy equation~\eqref{eq:partially_overlapping_graphs} will generically only give rise to three nontrivial scattering facets, and can thus give rise to intersections of the proper dimension. 

Let us pause here to highlight the essential role being played by the numerators of canonical forms in this argument. Namely, these numerators eliminate any contribution coming from the intersection of the hyperplanes $\mathcal{W}_{\mathfrak{g}_1}$ and $\mathcal{W}_{\mathfrak{g}_2}$ outside of the polytope by vanishing along these intersections~\cite{Arkani-Hamed:2014dca}.  Thus, while the intersection of any two $k$-dimensional hyperplanes in projective space has dimension $k-1$, the intersection of two facets of the cosmological polytope can have lower dimension as a result of these numerators.

This then is the mechanism by which the Steinmann relations are enforced by the structure of the cosmological polytope. Any two facets corresponding to partially-overlapping momentum channels intersect outside of the cosmological polytope (up to higher-codimension boundaries); thus, taking sequential residues on these facets yields zero due to the vanishing of the numerator along this intersection. The corresponding sequential cut---and sequential discontinuity---thereby also vanish.

As a final remark, it is important to note that the argument above holds for any graph and for any number of external states. In particular, it is well known that the double discontinuity of the box graph does not vanish upon analytic continuation outside the physical region~\cite{Stapp:1971hh}. However, there is no contradiction with the analysis just presented, as this nonzero double discontinuity is not visible in Lorentzian signature (and, consequently, with real energies), while the scattering facet (and indeed the whole cosmological polytope) is intrinsically Lorentzian.


\section{Steinmann relations and the wavefunction of the universe}\label{sec:ELI}

Let us now carry out the same analysis on the full cosmological polytope $\mathcal{P}_{\mathcal{G}}$. Namely, we ask whether the intersection of a pair of codimension-one faces $\mathcal{P}_{\mathcal{G}} \cap \mathcal{W}_{\mathfrak{g}_1}$ and $\mathcal{P}_{\mathcal{G}} \cap \mathcal{W}_{\mathfrak{g}_2}$ on the cosmological polytope corresponds to a codimension-two face when the subgraphs $\mathfrak{g}_1$ and $\mathfrak{g}_2$ satisfy equation~\eqref{eq:partially_overlapping_graphs}.

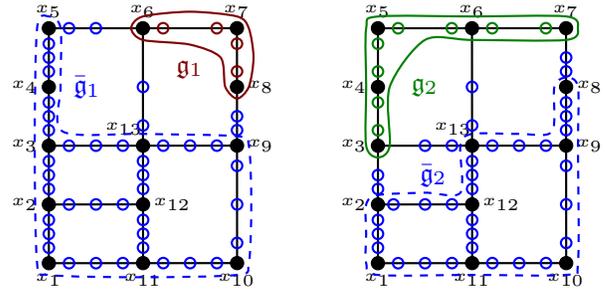
\begin{figure}
 \centering
 \vspace{-.5cm}
 \begin{tikzpicture}[ball/.style = {circle, draw, align=center, anchor=north, inner sep=0}, cross/.style={cross out, draw, minimum size=2*(#1-\pgflinewidth), inner sep=0pt, outer sep=0pt}, scale=1.25, transform shape]
  \begin{scope}
   \coordinate[label=left:{\tiny $x_3$}] (v1) at (0,0);
   \coordinate[label=above:{\tiny $x_5$}] (v2) at ($(v1)+(0,1.25)$);
   \coordinate[label=above:{\tiny $x_6$}] (v3) at ($(v2)+(1,0)$);
   \coordinate[label=above:{\tiny $x_7$}] (v4) at ($(v3)+(1,0)$);
   \coordinate[label=right:{\tiny $x_8$}] (v5) at ($(v4)-(0,.625)$);
   \coordinate[label=right:{\tiny $x_9$}] (v6) at ($(v5)-(0,.625)$);
   \coordinate[label={{[shift={(-0.2,0)}]\tiny $x_{13}$}}] (v7) at ($(v6)-(1,0)$);
   \coordinate[label=below:{\tiny $x_1$}] (v8) at ($(v1)-(0,1.25)$);
   \coordinate[label=below:{\tiny $x_{11}$}] (v9) at ($(v7)-(0,1.25)$);
   \coordinate[label=below:{\tiny $x_{10}$}] (v10) at ($(v6)-(0,1.25)$);
   \coordinate[label=left:{\tiny $x_{2}$}] (v11) at ($(v1)!0.5!(v8)$);
   \coordinate[label=right:{\tiny $x_{12}$}] (v12) at ($(v7)!0.5!(v9)$);
   \coordinate[label=left:{\tiny $x_{4}$}] (v13) at ($(v1)!0.5!(v2)$);   
   \draw[thick] (v2) -- (v4) -- (v10) -- (v8)  -- cycle;
   \draw[thick] (v1) -- (v6);   
   \draw[thick] (v3) -- (v9);
   \draw[thick] (v11) -- (v12);   
   \draw[fill=black] (v1) circle (2pt);
   \draw[fill=black] (v2) circle (2pt);
   \draw[fill=black] (v3) circle (2pt);
   \draw[fill=black] (v4) circle (2pt);
   \draw[fill=black] (v5) circle (2pt);
   \draw[fill=black] (v6) circle (2pt);
   \draw[fill=black] (v7) circle (2pt);
   \draw[fill=black] (v8) circle (2pt);
   \draw[fill=black] (v9) circle (2pt);
   \draw[fill=black] (v10) circle (2pt);
   \draw[fill=black] (v11) circle (2pt);
   \draw[fill=black] (v12) circle (2pt);
   \draw[fill=black] (v13) circle (2pt);
   \coordinate (v113) at ($(v1)!0.5!(v13)$);
   \coordinate (v132) at ($(v2)!0.5!(v13)$);
   \coordinate (v23) at ($(v2)!0.5!(v3)$);
   \coordinate (v34) at ($(v3)!0.5!(v4)$);
   \coordinate (v45) at ($(v4)!0.5!(v5)$);
   \coordinate (v56) at ($(v5)!0.5!(v6)$);
   \coordinate (v67) at ($(v6)!0.5!(v7)$);
   \coordinate (v71) at ($(v7)!0.5!(v1)$);
   \coordinate (v37) at ($(v3)!0.5!(v7)$);
   \coordinate (v111) at ($(v1)!0.5!(v11)$);
   \coordinate (v118) at ($(v11)!0.5!(v8)$);
   \coordinate (v712) at ($(v7)!0.5!(v12)$);
   \coordinate (v129) at ($(v12)!0.5!(v9)$);
   \coordinate (v610) at ($(v6)!0.5!(v10)$);
   \coordinate (v89) at ($(v8)!0.5!(v9)$);
   \coordinate (v910) at ($(v9)!0.5!(v10)$);
   \coordinate (v1112) at ($(v11)!0.5!(v12)$);
   \node[ball,text width=.18cm,thick,color=red!50!black,right=.15cm of v3, scale=.625] {};
   \node[ball,text width=.18cm,thick,color=red!50!black,left=.15cm of v4, scale=.625] {};
   \node[ball,text width=.18cm,thick,color=red!50!black,below=.1cm of v4, scale=.625] {};
   \node[ball,text width=.18cm,thick,color=red!50!black,above=.1cm of v5, scale=.625] {};
   \node[ball,text width=.18cm,thick,color=blue, anchor=center, scale=.625] at (v23) {};
   \node[ball,text width=.18cm,thick,color=blue, right=.15cm of v2, scale=.625] {};
   \node[ball,text width=.18cm,thick,color=blue, anchor=center, scale=.625] at (v37) {};
   \node[ball,text width=.18cm,thick,color=blue, above=.15cm of v7, scale=.625] {};
   \node[ball,text width=.18cm,thick,color=blue, anchor=center, scale=.625] at (v56) {};
   \node[ball,text width=.18cm,thick,color=blue, above=.1cm of v6, scale=.625] {};
   \node[ball,text width=.18cm,thick,color=blue, below=.1cm of v2, scale=.625] {};
   \node[ball,text width=.18cm,thick,color=blue, anchor=center, scale=.625] at (v132) {};
   \node[ball,text width=.18cm,thick,color=blue, above=.1cm of v13, scale=.625] {};
   \node[ball,text width=.18cm,thick,color=blue, below=.1cm of v13, scale=.625] {};
   \node[ball,text width=.18cm,thick,color=blue, anchor=center, scale=.625] at (v113) {};
   \node[ball,text width=.18cm,thick,color=blue, above=.1cm of v1, scale=.625] {};
   \node[ball,text width=.18cm,thick,color=blue, right=.15cm of v1, scale=.625] {};
   \node[ball,text width=.18cm,thick,color=blue, below=.1cm of v1, scale=.625] {};
   \node[ball,text width=.18cm,thick,color=blue, anchor=center, scale=.625] at (v111) {};
   \node[ball,text width=.18cm,thick,color=blue, above=.1cm of v11, scale=.625] {};
   \node[ball,text width=.18cm,thick,color=blue, right=.15cm of v11, scale=.625] {};
   \node[ball,text width=.18cm,thick,color=blue, below=.1cm of v11, scale=.625] {};
   \node[ball,text width=.18cm,thick,color=blue, anchor=center, scale=.625] at (v118) {};
   \node[ball,text width=.18cm,thick,color=blue, above=.1cm of v8, scale=.625] {};
   \node[ball,text width=.18cm,thick,color=blue, right=.15cm of v8, scale=.625] {};
   \node[ball,text width=.18cm,thick,color=blue, anchor=center, scale=.625] at (v89) {};
   \node[ball,text width=.18cm,thick,color=blue, left=.15cm of v9, scale=.625] {};
   \node[ball,text width=.18cm,thick,color=blue, above=.1cm of v9, scale=.625] {};
   \node[ball,text width=.18cm,thick,color=blue, right=.15cm of v9, scale=.625] {};
   \node[ball,text width=.18cm,thick,color=blue, anchor=center, scale=.625] at (v129) {};
   \node[ball,text width=.18cm,thick,color=blue, left=.15cm of v12, scale=.625] {};
   \node[ball,text width=.18cm,thick,color=blue, above=.1cm of v12, scale=.625] {};
   \node[ball,text width=.18cm,thick,color=blue, below=.1cm of v12, scale=.625] {};
   \node[ball,text width=.18cm,thick,color=blue, anchor=center, scale=.625] at (v712) {};
   \node[ball,text width=.18cm,thick,color=blue, left=.15cm of v7, scale=.625] {};
   \node[ball,text width=.18cm,thick,color=blue, right=.15cm of v7, scale=.625] {};
   \node[ball,text width=.18cm,thick,color=blue, below=.1cm of v7, scale=.625] {};
   \node[ball,text width=.18cm,thick,color=blue, anchor=center, scale=.625] at (v67) {};
   \node[ball,text width=.18cm,thick,color=blue, left=.15cm of v6, scale=.625] {};
   \node[ball,text width=.18cm,thick,color=blue, below=.15cm of v6, scale=.625] {};
   \node[ball,text width=.18cm,thick,color=blue, anchor=center, scale=.625] at (v610) {};
   \node[ball,text width=.18cm,thick,color=blue, left=.15cm of v10, scale=.625] {};
   \node[ball,text width=.18cm,thick,color=blue, above=.15cm of v10, scale=.625] {};
   \node[ball,text width=.18cm,thick,color=blue, anchor=center, scale=.625] at (v71) {};
   \node[ball,text width=.18cm,thick,color=blue, anchor=center, scale=.625] at (v1112) {};
   \node[ball,text width=.18cm,thick,color=blue, anchor=center, scale=.625] at (v910) {};
   \coordinate (a) at ($(v3)-(.125,0)$);
   \coordinate (b) at ($(v3)+(0,.125)$);
   \coordinate (c) at ($(v34)+(0,.175)$);
   \coordinate (d) at ($(v4)+(0,.125)$);
   \coordinate (e) at ($(v4)+(.125,0)$);
   \coordinate (f) at ($(v45)+(.175,0)$);
   \coordinate (g) at ($(v5)+(.125,0)$);
   \coordinate (h) at ($(v5)-(0,.125)$);
   \coordinate (i) at ($(v5)-(.125,0)$);
   \coordinate (j) at ($(v45)-(.175,0)$);
   \coordinate (k) at ($(v34)-(0,.175)$);
   \coordinate (l) at ($(v3)-(0,.125)$);
   \draw [thick, red!50!black] plot [smooth cycle] coordinates {(a) (b) (c) (d) (e) (f) (g) (h) (i) (j) (k) (l)};
   \node[below=.05cm of k, color=red!50!black] {\footnotesize $\displaystyle\mathfrak{g}_1$};
   \coordinate (a2) at ($(v2)+(.125,0)$);
   \coordinate (b2) at ($(v2)+(0,.125)$);
   \coordinate (c2) at ($(v2)-(.125,0)$);
   \coordinate (d2) at ($(v1)-(.15,0)$);
   \coordinate (e2) at ($(v8)-(.125,0)$);
   \coordinate (f2) at ($(v8)-(0,.125)$);
   \coordinate (g2) at ($(v9)-(0,.15)$);
   \coordinate (h2) at ($(v10)-(0,.125)$);
   \coordinate (i2) at ($(v10)+(.125,0)$);
   \coordinate (j2) at ($(v610)+(.15,0)$);
   \coordinate (k2) at ($(v6)+(.125,0)$);
   \coordinate (l2) at ($(v6)+(0,.1)$);
   \coordinate (m2) at ($(v7)+(0,.125)$);
   \coordinate (n2) at ($(v1)+(.15,.15)$);
   \coordinate (o2) at ($(v13)+(.125,0)$);
   \draw [thick, dashed, blue] plot [smooth cycle] coordinates {(a2) (b2) (c2) (d2) (e2) (f2) (g2) (h2) (i2) (j2) (k2) (l2) (m2) (n2) (o2)};
   \node[right=.15cm of v13, color=blue] {\footnotesize $\displaystyle\bar{\mathfrak{g}}_1$};
  \end{scope}
  \begin{scope}[shift={(3.5, 0)}, transform shape]
   \coordinate[label=left:{\tiny $x_3$}] (v1) at (0,0);
   \coordinate[label=above:{\tiny $x_5$}] (v2) at ($(v1)+(0,1.25)$);
   \coordinate[label=above:{\tiny $x_6$}] (v3) at ($(v2)+(1,0)$);
   \coordinate[label=above:{\tiny $x_7$}] (v4) at ($(v3)+(1,0)$);
   \coordinate[label=right:{\tiny $x_8$}] (v5) at ($(v4)-(0,.625)$);
   \coordinate[label=right:{\tiny $x_9$}] (v6) at ($(v5)-(0,.625)$);
   \coordinate[label={{[shift={(-0.2,0)}]\tiny $x_{13}$}}] (v7) at ($(v6)-(1,0)$);
   \coordinate[label=below:{\tiny $x_1$}] (v8) at ($(v1)-(0,1.25)$);
   \coordinate[label=below:{\tiny $x_{11}$}] (v9) at ($(v7)-(0,1.25)$);
   \coordinate[label=below:{\tiny $x_{10}$}] (v10) at ($(v6)-(0,1.25)$);
   \coordinate[label=left:{\tiny $x_{2}$}] (v11) at ($(v1)!0.5!(v8)$);
   \coordinate[label=right:{\tiny $x_{12}$}] (v12) at ($(v7)!0.5!(v9)$);
   \coordinate[label=left:{\tiny $x_{4}$}] (v13) at ($(v1)!0.5!(v2)$);   
   \draw[thick] (v2) -- (v4) -- (v10) -- (v8)  -- cycle;
   \draw[thick] (v1) -- (v6);   
   \draw[thick] (v3) -- (v9);
   \draw[thick] (v11) -- (v12);   
   \draw[fill=black] (v1) circle (2pt);
   \draw[fill=black] (v2) circle (2pt);
   \draw[fill=black] (v3) circle (2pt);
   \draw[fill=black] (v4) circle (2pt);
   \draw[fill=black] (v5) circle (2pt);
   \draw[fill=black] (v6) circle (2pt);
   \draw[fill=black] (v7) circle (2pt);
   \draw[fill=black] (v8) circle (2pt);
   \draw[fill=black] (v9) circle (2pt);
   \draw[fill=black] (v10) circle (2pt);
   \draw[fill=black] (v11) circle (2pt);
   \draw[fill=black] (v12) circle (2pt);
   \draw[fill=black] (v13) circle (2pt);
   \coordinate (v113) at ($(v1)!0.5!(v13)$);
   \coordinate (v132) at ($(v2)!0.5!(v13)$);
   \coordinate (v23) at ($(v2)!0.5!(v3)$);
   \coordinate (v34) at ($(v3)!0.5!(v4)$);
   \coordinate (v45) at ($(v4)!0.5!(v5)$);
   \coordinate (v56) at ($(v5)!0.5!(v6)$);
   \coordinate (v67) at ($(v6)!0.5!(v7)$);
   \coordinate (v71) at ($(v7)!0.5!(v1)$);
   \coordinate (v37) at ($(v3)!0.5!(v7)$);
   \coordinate (v111) at ($(v1)!0.5!(v11)$);
   \coordinate (v118) at ($(v11)!0.5!(v8)$);
   \coordinate (v712) at ($(v7)!0.5!(v12)$);
   \coordinate (v129) at ($(v12)!0.5!(v9)$);
   \coordinate (v610) at ($(v6)!0.5!(v10)$);
   \coordinate (v89) at ($(v8)!0.5!(v9)$);
   \coordinate (v910) at ($(v9)!0.5!(v10)$);
   \coordinate (v1112) at ($(v11)!0.5!(v12)$);
   \coordinate (v1r) at ($(v1)+(.15,0)$);
   \coordinate (v1d) at ($(v1)-(0,.15)$);
   \coordinate (v3d) at ($(v3)-(0,.15)$);
   \coordinate (v4d) at ($(v4)-(0,.15)$);
   \node[ball,text width=.18cm,thick,color=green!50!black,left=.15cm of v4, scale=.625] {};
   \node[ball,text width=.18cm,thick,color=green!50!black,right=.15cm of v3, scale=.625] {};
   \node[ball,text width=.18cm,thick,color=green!50!black,left=.15cm of v3, scale=.625] {};
   \node[ball,text width=.18cm,thick,color=green!50!black,right=.15cm of v2, scale=.625] {};
   \node[ball,text width=.18cm,thick,color=green!50!black,below=.1cm of v2, scale=.625] {};
   \node[ball,text width=.18cm,thick,color=green!50!black,above=.1cm of v13, scale=.625] {};
   \node[ball,text width=.18cm,thick,color=green!50!black,below=.1cm of v13, scale=.625] {};
   \node[ball,text width=.18cm,thick,color=green!50!black,above=.1cm of v1, scale=.625] {};   
   \node[ball,text width=.18cm,thick,color=blue, anchor=center, scale=.625] at (v71) {};
   \node[ball,text width=.18cm,thick,color=blue, left=.15cm of v7, scale=.625] {};
   \node[ball,text width=.18cm,thick,color=blue, above=.15cm of v7, scale=.625] {};
   \node[ball,text width=.18cm,thick,color=blue, anchor=center, scale=.625] at (v37) {};
   \node[ball,text width=.18cm,thick,color=blue, anchor=center, scale=.625] at (v45) {};
   \node[ball,text width=.18cm,thick,color=blue, above=.1cm of v5, scale=.625] {};   
   \node[ball,text width=.18cm,thick,color=blue, above=.1cm of v11, scale=.625] {};
   \node[ball,text width=.18cm,thick,color=blue, anchor=center, scale=.625] at (v111) {};
   \node[ball,text width=.18cm,thick,color=blue, below=.1cm of v5, scale=.625] {};
   \node[ball,text width=.18cm,thick,color=blue, anchor=center, scale=.625] at (v56) {};
   \node[ball,text width=.18cm,thick,color=blue, above=.1cm of v6, scale=.625] {};
   \node[ball,text width=.18cm,thick,color=blue, left=.15cm of v6, scale=.625] {};
   \node[ball,text width=.18cm,thick,color=blue, below=.15cm of v6, scale=.625] {};
   \node[ball,text width=.18cm,thick,color=blue, anchor=center, scale=.625] at (v610) {};
   \node[ball,text width=.18cm,thick,color=blue, above=.15cm of v10, scale=.625] {};
   \node[ball,text width=.18cm,thick,color=blue, left=.15cm of v10, scale=.625] {};
   \node[ball,text width=.18cm,thick,color=blue, anchor=center, scale=.625] at (v910) {};
   \node[ball,text width=.18cm,thick,color=blue, right=.15cm of v9, scale=.625] {};
   \node[ball,text width=.18cm,thick,color=blue, above=.1cm of v9, scale=.625] {};
   \node[ball,text width=.18cm,thick,color=blue, left=.15cm of v9, scale=.625] {};
   \node[ball,text width=.18cm,thick,color=blue, anchor=center, scale=.625] at (v129) {};
   \node[ball,text width=.18cm,thick,color=blue, anchor=center, scale=.625] at (v89) {};
   \node[ball,text width=.18cm,thick,color=blue, right=.15cm of v8, scale=.625] {};
   \node[ball,text width=.18cm,thick,color=blue, above=.1cm of v8, scale=.625] {};
   \node[ball,text width=.18cm,thick,color=blue, anchor=center, scale=.625] at (v118) {};
   \node[ball,text width=.18cm,thick,color=blue, below=.1cm of v11, scale=.625] {};
   \node[ball,text width=.18cm,thick,color=blue, right=.1cm of v11, scale=.625] {};
   \node[ball,text width=.18cm,thick,color=blue, anchor=center, scale=.625] at (v1112) {};
   \node[ball,text width=.18cm,thick,color=blue, left=.15cm of v12, scale=.625] {};
   \node[ball,text width=.18cm,thick,color=blue, below=.1cm of v12, scale=.625] {};
   \node[ball,text width=.18cm,thick,color=blue, above=.1cm of v12, scale=.625] {};
   \node[ball,text width=.18cm,thick,color=blue, anchor=center, scale=.625] at (v712) {};
   \node[ball,text width=.18cm,thick,color=blue, right=.15cm of v7, scale=.625] {};
   \node[ball,text width=.18cm,thick,color=blue, below=.1cm of v7, scale=.625] {};
   \node[ball,text width=.18cm,thick,color=blue, anchor=center, scale=.625] at (v67) {};
   \coordinate (a) at ($(v4)+(.125,0)$);
   \coordinate (b) at ($(v4)+(0,.125)$);
   \coordinate (c) at ($(v3)+(0,.125)$);
   \coordinate (d) at ($(v2)+(0,.125)$);
   \coordinate (e) at ($(v2)-(.125,0)$);
   \coordinate (f) at ($(v13)-(.125,0)$);
   \coordinate (g) at ($(v1)-(.125,0)$);
   \coordinate (gh) at ($(v1)-(0,.125)$);
   \coordinate (h) at ($(v1)+(.125,0)$);
   \coordinate (j) at ($(v13)+(.125,0)$);
   \coordinate (k) at ($(v23)-(0,.125)$);
   \coordinate (l) at ($(v3)-(0,.125)$);
   \coordinate (m) at ($(v4)-(0,.125)$);
   \draw [thick, green!50!black] plot [smooth cycle] coordinates {(a) (b) (c) (d) (e) (f) (g) (gh) (h) (j) (k) (l) (m)};
   \node[color=green!50!black] at ($(v13)+(.5,0)$) {\footnotesize $\displaystyle\mathfrak{g}_2$};
   \coordinate (a2) at ($(v5)+(.125,0)$);
   \coordinate (b2) at ($(v5)+(0,.1)$);
   \coordinate (c2) at ($(v5)-(.125,0)$);
   \coordinate (d2) at ($(v6)+(-.15,.15)$);
   \coordinate (e2) at ($(v67)+(0,.125)$);
   \coordinate (f2) at ($(v7)+(0,.125)$);
   \coordinate (g2) at ($(v7)-(.125,0)$);
   \coordinate (h2) at ($(v12)+(-.125,.125)$);
   \coordinate (i2) at ($(v1112)+(0,.125)$);
   \coordinate (j2) at ($(v11)+(0,.1)$);
   \coordinate (k2) at ($(v11)-(.125,0)$);
   \coordinate (l2) at ($(v118)-(.125,0)$);
   \coordinate (m2) at ($(v8)-(.125,0)$);
   \coordinate (n2) at ($(v8)-(0,.125)$);
   \coordinate (o2) at ($(v9)-(0,.125)$);
   \coordinate (p2) at ($(v10)-(0,.125)$);
   \coordinate (q2) at ($(v10)+(.125,0)$);
   \coordinate (r2) at ($(v6)+(.125,0)$);
   \draw [thick, dashed, blue] plot [smooth cycle] coordinates {(a2) (b2) (c2) (d2) (e2) (f2) (g2) (h2) (i2) (j2) (k2) (l2) (m2) (n2) (o2) (p2) (q2) (r2)};
   \node[left=.15cm of v712, color=blue] {\footnotesize $\displaystyle\bar{\mathfrak{g}}_2$};
  \end{scope}
 \end{tikzpicture}
 \caption{A pair of partially-overlapping codimension-one faces $\mathcal{P}_{\mathcal{G}}\cap\mathcal{W}_{\mathfrak{g}_1}$ and $\mathcal{P}_{\mathcal{G}}\cap\mathcal{W}_{\mathfrak{g}_2}$ of the cosmological polytope $\mathcal{P}_{\mathcal{G}}$.
 The markings $\bluecircle$ depict the complementary subgraphs $\bar{\mathfrak{g}}_j$ as well as the vertices on the cut edges. Together, these vertices form the polytope $\mathcal{P}_{\mathfrak{g}_j\cup\centernot{\mathcal{E}}}$.}
 \label{fig:facetcp}
\end{figure}

On each of the codimension-one faces $\mathcal{P}_{\mathcal{G}} \cap \mathcal{W}_{\mathfrak{g}_j}$, the canonical function factorizes into a pair of lower-dimension polytopes. Namely,
\begin{equation}\label{eq:WFfact}
 \Omega(\mathcal{Y},\,\mathcal{P}_{\mathcal{G}} \cap \mathcal{W}_{\mathfrak{g}_j})\:=\:\Omega(\mathcal{Y}_{\mathfrak{g}_j},\,\mathcal{S}_{\mathfrak{g}_j})\times 
 \Omega(\mathcal{Y}_{\bar{\mathfrak{g}}_j\cup\centernot{\mathcal{E}}},\,\mathcal{P}_{\bar{\mathfrak{g}}_j\cup\centernot{\mathcal{E}}})
\end{equation}
where $\Omega(\mathcal{Y}_{\mathfrak{g}_j},\,\mathcal{S}_{\mathfrak{g}_j})$ is the canonical function of the scattering facet $\mathcal{S}_{\mathfrak{g}_j}$, and  $\Omega(\mathcal{Y}_{\bar{\mathfrak{g}}_j\cup\centernot{\mathcal{E}}},\,\mathcal{P}_{\bar{\mathfrak{g}}_j\cup\centernot{\mathcal{E}}})$ is the canonical function of the polytope $\mathcal{P}_{\bar{\mathfrak{g}}_j\cup\centernot{\mathcal{E}}}$, defined to be the convex hull of the vertices associated to the complementary graph $\bar{\mathfrak{g}}_j$ and the cut edges $\centernot{\mathcal{E}}$.\footnote{Note that we're abusing notation here, insofar as \(\bar{\mathfrak{g}}_j \cup \centernot{\mathcal{E}}\) is not a graph in the ordinary sense; we include in it the edges in \(\centernot{\mathcal{E}}\), but not the sites outside of \(\bar{\mathfrak{g}}_j\) that these edges are incident with.} This factorization is depicted in Figure~\ref{fig:facetcp}. 

In more detail, the canonical function $\Omega(\mathcal{Y}_{\bar{\mathfrak{g}}_j\cup\centernot{\mathcal{E}}},\,\mathcal{P}_{\bar{\mathfrak{g}}_j\cup\centernot{\mathcal{E}}})$ is related to the lower-point wavefunction $\psi_{\bar{\mathfrak{g}}_j}$ associated with the subgraph $\bar{\mathfrak{g}}_j$ \cite{Arkani-Hamed:2017fdk, Benincasa:2018ssx}. Specifically, it is given by a sum over the positive and negative energy solutions for the energy on the cut edges $\centernot{\mathcal{E}}$, divided by the
product of the energies associated with the cut edges \cite{Baumann:2020dch}:
\begin{equation}\label{eq:canonical_func_wf}
 \begin{split}
  \Omega(\mathcal{Y}_{\bar{\mathfrak{g}}_j\cup\centernot{\mathcal{E}}},\,\mathcal{P}_{\bar{\mathfrak{g}}_j\cup\centernot{\mathcal{E}}})\:&=\: \sum_{\{\sigma_e\}=\pm 1} \frac{\psi_{\bar{\mathfrak{g}}_j}
   \left(x_{v}(\sigma_e),y_{e}\right)}{\prod_{e\in\centernot{\mathcal{E}}} 2y_e}
 \end{split}
\end{equation}
where, like in equation~\eqref{eq:WFint}, the arguments of $\psi_{\bar{\mathfrak{g}}_j}$ are the energies associated with the sites and edges in $\bar{\mathfrak{g}}_j$. However, the site energies have been shifted by the energies of the cut edges $\centernot{\mathcal{E}}$, namely
\begin{equation} \label{eq:shifted_energies}
    x_{v}(\sigma_e ) = x_{v} + \sum_{e \in \centernot{\mathcal{E}} \cap \mathcal{E}_{v}} \sigma_e y_e \, ,
\end{equation}
where $\mathcal{E}_v$ denotes the set of edges that depart from the site $v$ (thus, the energies associated with sites that aren't connected to a cut edge remain unshifted).  
Importantly, the explicit factors of $(2y_e)^{-1}$ in \eqref{eq:shifted_energies} cancel after carrying out the sum, and therefore do not correspond to real poles~\cite{Benincasa:2018ssx}.

\begin{figure}
 \centering
 \vspace{-.5cm}
 \begin{tikzpicture}[ball/.style = {circle, draw, align=center, anchor=north, inner sep=0}, cross/.style={cross out, draw, minimum size=2*(#1-\pgflinewidth), inner sep=0pt, outer sep=0pt}, scale=1.25, transform shape]
  \begin{scope}
   \coordinate[label=left:{\tiny $x_3$}] (v1) at (0,0);
   \coordinate[label=above:{\tiny $x_5$}] (v2) at ($(v1)+(0,1.25)$);
   \coordinate[label=above:{\tiny $x_6$}] (v3) at ($(v2)+(1,0)$);
   \coordinate[label=above:{\tiny $x_7$}] (v4) at ($(v3)+(1,0)$);
   \coordinate[label=right:{\tiny $x_8$}] (v5) at ($(v4)-(0,.625)$);
   \coordinate[label=right:{\tiny $x_9$}] (v6) at ($(v5)-(0,.625)$);
   \coordinate[label={{[shift={(-0.2,0)}]\tiny $x_{13}$}}] (v7) at ($(v6)-(1,0)$);
   \coordinate[label=below:{\tiny $x_1$}] (v8) at ($(v1)-(0,1.25)$);
   \coordinate[label=below:{\tiny $x_{11}$}] (v9) at ($(v7)-(0,1.25)$);
   \coordinate[label=below:{\tiny $x_{10}$}] (v10) at ($(v6)-(0,1.25)$);
   \coordinate[label=left:{\tiny $x_{2}$}] (v11) at ($(v1)!0.5!(v8)$);
   \coordinate[label=right:{\tiny $x_{12}$}] (v12) at ($(v7)!0.5!(v9)$);
   \coordinate[label=left:{\tiny $x_{4}$}] (v13) at ($(v1)!0.5!(v2)$);   
   \draw[thick] (v2) -- (v4) -- (v10) -- (v8)  -- cycle;
   \draw[thick] (v1) -- (v6);   
   \draw[thick] (v3) -- (v9);
   \draw[thick] (v11) -- (v12);   
   \draw[fill=black] (v1) circle (2pt);
   \draw[fill=black] (v2) circle (2pt);
   \draw[fill=black] (v3) circle (2pt);
   \draw[fill=black] (v4) circle (2pt);
   \draw[fill=black] (v5) circle (2pt);
   \draw[fill=black] (v6) circle (2pt);
   \draw[fill=black] (v7) circle (2pt);
   \draw[fill=black] (v8) circle (2pt);
   \draw[fill=black] (v9) circle (2pt);
   \draw[fill=black] (v10) circle (2pt);
   \draw[fill=black] (v11) circle (2pt);
   \draw[fill=black] (v12) circle (2pt);
   \draw[fill=black] (v13) circle (2pt);
   \coordinate (v113) at ($(v1)!0.5!(v13)$);
   \coordinate (v132) at ($(v2)!0.5!(v13)$);
   \coordinate (v23) at ($(v2)!0.5!(v3)$);
   \coordinate (v34) at ($(v3)!0.5!(v4)$);
   \coordinate (v45) at ($(v4)!0.5!(v5)$);
   \coordinate (v56) at ($(v5)!0.5!(v6)$);
   \coordinate (v67) at ($(v6)!0.5!(v7)$);
   \coordinate (v71) at ($(v7)!0.5!(v1)$);
   \coordinate (v37) at ($(v3)!0.5!(v7)$);
   \coordinate (v111) at ($(v1)!0.5!(v11)$);
   \coordinate (v118) at ($(v11)!0.5!(v8)$);
   \coordinate (v712) at ($(v7)!0.5!(v12)$);
   \coordinate (v129) at ($(v12)!0.5!(v9)$);
   \coordinate (v610) at ($(v6)!0.5!(v10)$);
   \coordinate (v89) at ($(v8)!0.5!(v9)$);
   \coordinate (v910) at ($(v9)!0.5!(v10)$);
   \coordinate (v1112) at ($(v11)!0.5!(v12)$);
   \coordinate (v1r) at ($(v1)+(.15,0)$);
   \coordinate (v1d) at ($(v1)-(0,.15)$);
   \coordinate (v3d) at ($(v3)-(0,.15)$);
   \coordinate (v3l) at ($(v3)-(.15,0)$);
   \coordinate (v4d) at ($(v4)-(0,.15)$);
   \coordinate (v5b) at ($(v5)-(0,.15)$);
   \coordinate (a) at ($(v3)-(.125,0)$);
   \coordinate (b) at ($(v3)+(0,.125)$);
   \coordinate (c) at ($(v34)+(0,.175)$);
   \coordinate (d) at ($(v4)+(0,.125)$);
   \coordinate (e) at ($(v4)+(.125,0)$);
   \coordinate (f) at ($(v4)-(0,.125)$);
   \coordinate (k) at ($(v34)-(0,.175)$);
   \coordinate (l) at ($(v3)-(0,.125)$);
   \draw [thick, red!50!black] (v5) circle (3pt);
   \node[color=red!50!black, scale=.75] at ($(v5)-(.475,0)$) {\footnotesize $\displaystyle\mathfrak{g}_1\cap\bar{\mathfrak{g}}_2$};   
   \draw [thick, red!50!green] plot [smooth cycle] coordinates {(a) (b) (c) (d) (e) (f) (k) (l)};
   \node[color=red!50!green, scale=.75] at ($(v34)+(0,.375)$) {\footnotesize $\displaystyle\mathfrak{g}_1\cap\mathfrak{g}_2$};   
   \coordinate (a2) at ($(v2)+(.125,0)$);
   \coordinate (d2) at ($(v2)+(0,.125)$);
   \coordinate (e2) at ($(v2)-(.125,0)$);
   \coordinate (f2) at ($(v13)-(.125,0)$);
   \coordinate (g2) at ($(v1)-(.125,0)$);
   \coordinate (gh2) at ($(v1)-(0,.125)$);
   \coordinate (h2) at ($(v1)+(.125,0)$);
   \coordinate (j2) at ($(v13)+(.125,0)$);
   \draw [thick, green!50!black] plot [smooth cycle] coordinates {(a2) (d2) (e2) (f2) (g2) (gh2) (h2) (j2)};
   \node[color=green!50!black, scale=.75] at ($(v13)+(.475,0)$) {\footnotesize $\displaystyle\mathfrak{g}_2\cap\bar{\mathfrak{g}}_1$};   
   \node[ball,text width=.18cm,thick,color=red!50!green,right=.15cm of v3, scale=.625] {};
   \node[ball,text width=.18cm,thick,color=red!50!green,left=.15cm of v4, scale=.625] {};
   \node[ball,text width=.18cm,thick,color=red,right=.15cm of v2, scale=.625] {};
   \node[ball,text width=.18cm,thick,color=green!50!black,below=.075cm of v2, scale=.625] {};
   \node[ball,text width=.18cm,thick,color=green!50!black,above=.075cm of v13, scale=.625] {};
   \node[ball,text width=.18cm,thick,color=green!50!black,below=.075cm of v13, scale=.625] {};
   \node[ball,text width=.18cm,thick,color=green!50!black,above=.075cm of v1, scale=.625] {};
   \node[ball,text width=.18cm,thick,color=red,above=.075cm of v5, scale=.625] {};
   %
   \node[ball,text width=.18cm,thick,color=blue,above=.1cm of v11, scale=.625] {};
   \node[ball,text width=.18cm,thick,color=blue, anchor=center, scale=.625] at (v111) {};
   \node[ball,text width=.18cm,thick,color=blue,above=.15cm of v7, scale=.625] {};
   \node[ball,text width=.18cm,thick,color=blue, anchor=center, scale=.625] at (v37) {};
   \node[ball,text width=.18cm,thick,color=blue,left=.15cm of v7, scale=.625] {};
   \node[ball,text width=.18cm,thick,color=blue, anchor=center, scale=.625] at (v71) {};
   \node[ball,text width=.18cm,thick,color=blue,above=.1cm of v6, scale=.625] {};
   \node[ball,text width=.18cm,thick,color=blue, anchor=center, scale=.625] at (v56) {}; 
   \node[ball,text width=.18cm,thick,color=blue,below=.075cm of v7, scale=.625] {};
   \node[ball,text width=.18cm,thick,color=blue,right=.15cm of v7, scale=.625] {};
   \node[ball,text width=.18cm,thick,color=blue, anchor=center, scale=.625] at (v67) {};
   \node[ball,text width=.18cm,thick,color=blue,left=.15cm of v6, scale=.625] {};
   \node[ball,text width=.18cm,thick,color=blue,below=.15cm of v6, scale=.625] {};
   \node[ball,text width=.18cm,thick,color=blue, anchor=center, scale=.625] at (v712) {};
   \node[ball,text width=.18cm,thick,color=blue,above=.075cm of v12, scale=.625] {};
   \node[ball,text width=.18cm,thick,color=blue,left=.15cm of v12, scale=.625] {};
   \node[ball,text width=.18cm,thick,color=blue,below=.075cm of v12, scale=.625] {};
   \node[ball,text width=.18cm,thick,color=blue, anchor=center, scale=.625] at (v1112) {};
   \node[ball,text width=.18cm,thick,color=blue,right=.15cm of v11, scale=.625] {};
   \node[ball,text width=.18cm,thick,color=blue,below=.075cm of v11, scale=.625] {};
   \node[ball,text width=.18cm,thick,color=blue, anchor=center, scale=.625] at (v118) {};
   \node[ball,text width=.18cm,thick,color=blue,above=.075cm of v8, scale=.625] {};
   \node[ball,text width=.18cm,thick,color=blue,right=.15cm of v8, scale=.625] {};
   \node[ball,text width=.18cm,thick,color=blue, anchor=center, scale=.625] at (v89) {};
   \node[ball,text width=.18cm,thick,color=blue,left=.15cm of v9, scale=.625] {};
   \node[ball,text width=.18cm,thick,color=blue,above=.075cm of v9, scale=.625] {};
   \node[ball,text width=.18cm,thick,color=blue,right=.15cm of v9, scale=.625] {};
   \node[ball,text width=.18cm,thick,color=blue, anchor=center, scale=.625] at (v910) {};
   \node[ball,text width=.18cm,thick,color=blue,left=.15cm of v10, scale=.625] {};
   \node[ball,text width=.18cm,thick,color=blue,above=.15cm of v10, scale=.625] {};
   \node[ball,text width=.18cm,thick,color=blue, anchor=center, scale=.625] at (v610) {};
   \node[ball,text width=.18cm,thick,color=blue, anchor=center, scale=.625] at (v129) {};
   \coordinate (b3) at ($(v6)+(.15,0)$);
   \coordinate (c3) at ($(v6)+(0,.1)$);
   \coordinate (d3) at ($(v67)+(0,.1)$);
   \coordinate (e3) at ($(v7)+(0,.1)$);
   \coordinate (f3) at ($(v7)-(.1,0)$);
   \coordinate (g3) at ($(v712)-(.1,0)$);
   \coordinate (h3) at ($(v12)+(-.15,.15)$);
   \coordinate (i3) at ($(v1112)+(0,.15)$);
   \coordinate (j3) at ($(v11)+(0,.1)$);
   \coordinate (k3) at ($(v11)-(.15,0)$);
   \coordinate (l3) at ($(v118)-(.15,0)$);
   \coordinate (m3) at ($(v8)-(.15,0)$);
   \coordinate (n3) at ($(v8)-(0,.15)$);
   \coordinate (o3) at ($(v89)-(0,.15)$);
   \coordinate (p3) at ($(v9)-(0,.15)$);
   \coordinate (q3) at ($(v910)-(0,.15)$);
   \coordinate (r3) at ($(v10)-(0,.15)$);
   \coordinate (s3) at ($(v10)+(.15,0)$);
   \coordinate (t3) at ($(v610)+(.15,0)$);
   \draw [thick, blue] plot [smooth cycle] coordinates {(b3) (c3) (d3) (e3) (f3) (g3) (h3) (i3) (j3) (k3) (l3) (m3) (n3) (o3) (p3) (q3) (r3) (s3) (t3)};
   \node[color=blue, scale=.75] at ($(v712)-(.5,0)$) {\footnotesize $\displaystyle\bar{\mathfrak{g}}_1\cap\bar{\mathfrak{g}}_2$};   
  \end{scope}
 \end{tikzpicture}
 \caption{The intersection of the pair of facets depicted in Figure \ref{fig:facetcp}. This intersection  factorizes into a product of three lower-dimensional scattering facets associated with the graphs $\mathfrak{g}_1\cap\mathfrak{g}_2$, $\mathfrak{g}_1 \cap \bar{\mathfrak{g}}_2$, and $\mathfrak{g}_2 \cap \bar{\mathfrak{g}}_1$, and a lower-dimensional polytope associated with the graph $(\bar{\mathfrak{g}}_1\cap\bar{\mathfrak{g}}_2)\, \cup\! \centernot{\mathcal{E}}$.}
 \label{fig:CPfcod2}
\end{figure}
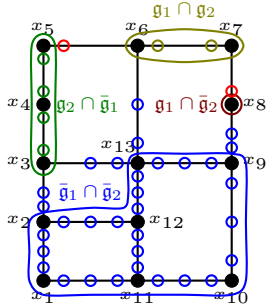

Now let us consider the intersection $\mathcal{P}_{\mathcal{G}} \cap \mathcal{W}_{\mathfrak{g}_1} \cap \mathcal{W}_{\mathfrak{g}_2}$ of a pair of codimension-one faces associated with the subgraphs $\mathfrak{g}_1$ and $\mathfrak{g}_2$. The canonical function on this intersection will in general factorize into three lower-point scattering facets associated with the graphs $\mathfrak{g}_1\cap\mathfrak{g}_2$, $\mathfrak{g}_1 \cap \bar{\mathfrak{g}}_2$, and $\mathfrak{g}_2 \cap \bar{\mathfrak{g}}_1$, and a lower-dimensional polytope associated with $\mathfrak{g}_c = (\bar{\mathfrak{g}}_1\cap\bar{\mathfrak{g}}_2)\,\cup\!\centernot{\mathcal{E}}$ similar to the second factor in equation~\eqref{eq:WFfact}. An example is shown in Figure~\ref{fig:CPfcod2}. There will also be a contribution coming from $\Sigma_{\centernot{\mathcal{E}}}$, which is
formed by the vertices on the cut edges among $\mathfrak{g}_1\cap\mathfrak{g}_2$, $\mathfrak{g}_1 \cap \bar{\mathfrak{g}}_2$, and $\mathfrak{g}_2 \cap \bar{\mathfrak{g}}_1$.

Like on the scattering facet, it is not sufficient to check that this intersection has enough vertices to (in principle) span a space of dimension $n_v+n_e-3$.
Thus, we again count the dimension of each of the polytopes that the canonical function factorizes into. As before, each scattering facet has dimension $n^{\mathfrak{g}}_{v}+n^{\mathfrak{g}}_{e}-2$, where $n^{\mathfrak{g}}_{v}$ and $n^{\mathfrak{g}}_{e}$ denote the number of sites and edges in the corresponding reduced graph $\mathfrak{g}$. The polytope associated with $\mathfrak{g}_c$ will have dimension 
$n^{\mathfrak{g}_c}_{v}+n^{\mathfrak{g}_c}_{e}+n_{\centernot{\bar{\mathcal{E}}}}-1$, where $n_{\centernot{\bar{\mathcal{E}}}}$ is the number of cut edges departing from this subgraph. Finally, the simplex $\Sigma_{\centernot{\mathcal{E}}}$ will have dimension $n_{\centernot{\mathcal{E}}}-1$, where $n_{\centernot{\mathcal{E}}}$ denotes the number of cut edges in $\mathcal{G}$ that contribute to this simplex. We thus find
\begin{equation}\label{eq:WFcount}
 \begin{split}
 \!\!\! \mbox{dim}\left(\mathcal{P}_{\mathcal{G}}\cap\mathcal{W}_{\mathfrak{g}_1}\cap\mathcal{W}_{\mathfrak{g}_2}\right)\:=\:
   &\sum_{\mathcal{S}_{\mathfrak{g}}}(n_v^{\mathfrak{g}}+n_e^{\mathfrak{g}}-1)+n_{\centernot{\mathcal{E}}}+\\
   &+(n^{\mathfrak{g}_c}_{v}+n^{\mathfrak{g}_c}_{e}+n_{\centernot{\bar{\mathcal{E}}}})-1,
 \end{split}
\end{equation}
where the sum is over the scattering facets related to the subgraphs $\mathfrak{g}_1\cap\mathfrak{g}_2$, $\mathfrak{g}_1\cap\bar{\mathfrak{g}}_2$ and $\mathfrak{g}_2\cap\bar{\mathfrak{g}}_1$. As each of the sites and edges in $\mathcal{G}$ only contribute to one of these factors, this
straightforwardly reduces to
\begin{equation}\label{eq:WFcount2}
  \mbox{dim}\left(\mathcal{P}_{\mathcal{G}}\cap\mathcal{W}_{\mathfrak{g}_1}\cap\mathcal{W}_{\mathfrak{g}_2}\right)
  \:=\:n_v+n_e-1-\sum_{\mathcal{S}_{\mathfrak{g}}} 1  \, .
\end{equation}
Thus, when $\mathfrak{g}_1$ and $\mathfrak{g}_2$ correspond to partially-overlapping channels and satisfy equation~\eqref{eq:partially_overlapping_graphs}, we see that this intersection has dimension $n_v + v_e - 4$, and the corresponding sequential cut vanishes. This implies new Steinmann-like relations for the wavefunction of the universe; in particular, we have
\begin{equation} \label{eq:steinmann_for_wfn}
 \! \mbox{Res}_{E_{\mathfrak{g}_1}}\left(\mbox{Res}_{E_{\mathfrak{g}_2}}\psi_{\mathcal{G}}\right)\:=\:0 \, , \quad\! \! \text{where } \begin{cases}\mathfrak{g}_1\nsubseteq\mathfrak{g}_2\\ \mathfrak{g}_2\nsubseteq \mathfrak{g}_1 \\
 \mathfrak{g}_1 \cap\mathfrak{g}_2 \neq\varnothing\end{cases} \!\!\!\!,
\end{equation}
where the energies $E_{\mathfrak{g}_j}$ were given in equation~\eqref{eq:subraph_energy}.
At tree level, where the integrals over site energies are known to give rise to polylogarithms~\cite{Arkani-Hamed:2017fdk,Hillman:2019wgh}, this can immediately be promoted to a restriction on the double discontinuities of $\Psi_{\mathcal{G}}$.\footnote{However, we note that (just like in for the standard Steinmann relations) one has to be careful how one chooses the analytic continuations used to actually compute these double discontinuities~\cite{Bourjaily:2020wvq}.}


\section{Conclusion and Outlook}\label{sec:CO}

Causality and unitarity constitute two of the basic pillars of our understanding of physical processes. In this work, we have explored how these properties emerge in flat-space from the structure of the wavefunction of the universe by studying its formulation in terms of cosmological polytopes. In particular, we have seen that the Steinmann relations are transparently encoded in the combinatorial properties of the canonical form of the scattering facet. Additionally, we have derived novel relations that directly restrict the analytic properties of the wavefunction of the universe. 

While the Steinmann relations for scattering amplitudes are understood to be implied by causality~\cite{Steinmann:1960soa, Steinmann:1960sob, Araki:1961hb, Ruelle:1961rd, Stapp:1971hh, Cahill:1973px, Lassalle:1974jm, Cahill:1973qp}, it is not yet clear whether this is true for the more general constraints we have derived. It would therefore be interesting to see these constraints follow directly from basic physical principles. In this respect, it is worth highlighting that the perturbative integrands of in--in correlation functions involving the operator \(\mathcal{O}\) can be written as \(R\)-products between interaction Hamiltonians and \(\mathcal{O}\). \(R\)-products were introduced to study causality~\cite{Lehmann:1957zz}, and play an important role in the formulation of the original Steinmann relations for correlation functions.

As the new relations we have derived constrain the functional form of the wavefunction of the universe, they may prove useful for bootstrapping this quantity (even at tree level). In particular, it is worth stressing that our analysis is valid in {\it any} FRW cosmology, as the object of our investigation was the universal integrand in equation~\eqref{eq:WFint}. The wavefunction for a specific background is obtained by integrating this quantity over external energies with appropriate coefficients $\tilde{\lambda}(\epsilon)$.\footnote{Moreover, the wavefunction we have studied here determines wavefunctions involving more general scalar states, which can be thought of as having a time-dependent mass; these latter wavefunctions are determined by recursion relations which the former wavefunctions enter as seeds~\cite{Benincasa:2019vqr}. Conformally-coupled scalars also play the role of seeds for correlation functions involving both scalars with more general conformal dimension, and states with spin, by applying suitable differential operators on them~\cite{Baumann:2019oyu, Baumann:2020dch}.}
Consequently, these constraints should provide valid input for the recently-inaugurated {\it boostless bootstrap program}~\cite{Pajer:2020wnj}.

Eventually, one would also like to find a generalization of cosmological polytopes that combinatorially encode the perturbative wavefunction of the universe in a single object, rather than just via the graphs that contribute to it. However, as the Steinmann-type constraints we have derived here apply graph-by-graph, these generalized cosmological polytopes would have to obey similar constraints.

Finally, while our analysis is fairly general, we highlight that it remains unclear whether these Steinmann-type relations for the wavefunction of the universe extend to processes involving states that do not have counterparts in flat space, which aren't entirely captured by the cosmological polytope.
We leave this important open question for future work.

\acknowledgments

\paragraph{Acknowledgments}--  The authors are supported in part by a grant from the Villum Fonden (No.~15369) and an ERC Starting Grant (No.~757978). P.B~ is also supported by the Danish National Research Foundation (DNRF91), and, in the last stage of this work, by the ``Atracci{\'o}n de Talento'' program of the Comunidad de Madrid under the grant 2019-T1/TIC -15568. A.J.M.~is also supported by a Carlsberg Postdoctoral Fellowship (CF18-0641). 

\bibliographystyle{utphys}
\bibliography{cprefs}

\end{document}